\documentclass[final,5p]{elsarticle}

\usepackage{amssymb}
\usepackage{mathtools}
\usepackage{lscape}

\usepackage{natbib}
\biboptions{comma,round,authoryear}

\usepackage{bm}

\usepackage{graphicx}
\usepackage{txfonts}
\usepackage{gensymb}
\usepackage{adjustbox}
\usepackage{multirow}
\usepackage{epstopdf}
\usepackage{tabulary}
\usepackage{tabularx}
\usepackage{hhline}
\usepackage{booktabs}
\usepackage{array}
\usepackage{subcaption}

\journal{Progress in Earth and Planetary Science}

\begin{document}

\begin{frontmatter}

\title{Regolith behavior under asteroid-level gravity conditions: low-velocity impact experiments}


\author[label1]{Julie Brisset}
\author[label1]{Joshua E. Colwell}
\author[label1]{Adrienne Dove}
\author[label1]{Sumayya Abukhalil}
\author[label1]{Christopher Cox}
\author[label1]{Nadia Mohammed}

\address[label1]{Center of Microgravity Research, University of Central Florida, 4111 Libra Drive, Orlando FL-32816}

\begin{abstract}

The dusty regolith covering the surfaces of asteroids and planetary satellites differs in size, shape, and composition from terrestrial soil particles and is subject to environmental conditions very different from those found on Earth. This regolith evolves in a low ambient pressure and low-gravity environment. Its response to low-velocity impacts, such as those that may accompany human and robotic exploration activities, may be completely different than what is encountered on Earth. Experimental studies of the response of planetary regolith in the relevant environmental conditions are thus necessary to facilitate future Solar System exploration activities. 

We combined the results and provided new data analysis elements for a series of impact experiments into simulated planetary regolith in low-gravity conditions using two experimental setups and a range of microgravity platforms. The Physics of Regolith Impacts in Microgravity Experiment (PRIME) flew on several parabolic aircraft flights, enabling the recording of impacts into granular materials at speeds of $\sim$4-230 cm/s. The COLLisions Into Dust Experiment (COLLIDE) is conceptually close to the PRIME setup. It flew on the Space Shuttle in 1998 and 2001 and more recently on the Blue Origin New Shepard rocket, recording impacts into simulated regolith at speeds between 1 and 120 cm/s.

Results of these experimental campaigns found that there is a significant change in the regolith behavior with the gravity environment. In a 10$^{-2}g$ environment (with $g$ being the gravity acceleration at the surface of the Earth), only embedding of the impactor was observed and ejecta production was produced for most impacts at $>$ 20 cm/s. Once at microgravity levels ($<10^{-4}g$), the lowest impact energies also produced impactor rebound. In these microgravity conditions, ejecta started to be produced for impacts at $>$ 10 cm/s. The measured ejecta speeds were somewhat lower than the ones measured at reduced-gravity levels, but the ejected masses were higher. In general, the mean ejecta velocity shows a power-law dependence on the impact energy with an index of $\sim$0.5. When projectile rebound occured, we observed that its coefficients of restitution on the bed of regolith simulant decrease by a factor of 10 with increasing impact speeds from $\sim$5 cm/s up to 100 cm/s. We could also observe an increased cohesion between the JSC-1 grains compared to the quartz sand targets.

\end{abstract}


\end{frontmatter}

\section{Introduction}
\label{s:intro}

Small airless bodies of the Solar System are known to be covered in a layer of regolith, composed of grains of varying sizes. Created by the bombardment of micrometeoroids or thermal disintegration, micron-sized and larger dust particles collect on the surfaces of asteroids, comets, and moons \citep{housen1979,housen1982}. Any space mission probing or landing on the surfaces of these small bodies has to interact with this regolith, at the very low surface gravity levels induced by the body's small mass. The first soft landings on such small bodies were performed by the NEAR-Shoemaker spacecraft on the asteroid Eros \citep{veverka2001}, Hayabusa on Itokawa \citep{kawaguchi2008}, and Philae on the comet 67P/Churyumov-Gerasimenko \citep{biele2015}. In the latter instances, the landing resulted in an unexpected response of the surface to the low-velocity impact of the lander, in particular the remarkable recurrent bouncing of Philae on the comet's surface after a first touch-down at around 1 m/s. 

Currently, two missions are en-route to asteroids: Hayabusa-2 carrying the MASCOT lander \citep{tsuda2013, jaumann2016}, due to arrive at Ryugu in 2018, and the OSIRIS-REx mission arriving at Bennu in 2018, planning to sample its surface using a Touch-And-Go mechanism \citep{lauretta2017}. In addition, the increased interest in In-Situ Resource Utilization (ISRU) and mining prospects will soon lead to more reconnaissance and pathfinder missions to small bodies of the Solar System. Understanding interactions with surface regolith at very low gravity levels has therefore become an imperative. Impacts at a few m/s down to a few cm/s into regolith layers are of particular interest to these current and future missions.

Motivated by these upcoming missions, theoretical and numerical work has recently been performed in order to investigate the behavior of granular material at the surface of small asteroids, such as Ryugu and Bennu \citep{scheeres2010, sanchez2014, hirabayashi2015, thuillet2017}. In particular, the response of a layer of regolith to a low-velocity impact is investigated in e.g. \cite{maurel2017} for MASCOT and \cite{sanchez2013}. Experimental work supporting these efforts are scarce and often performed under conditions very different from what can be expected during a landing/contact at the surface of small bodies (1$g$, hyper-velocity impacts). For example, the experiment campaign performed by \cite{housen2003} to investigate crater formation in porous materials studied impacts at $>$2 km/s \citep[see][ for a review of other high-speed impact experiments in porous materials]{holsapple2002, housen2011}. In order to investigate the relative influence of gravity on the impact cratering and ejecta production processes in very large impact events, \cite{housen1999} perfomed high-speed impacts into porous materials in a centrifuge at gravity levels up to 500$g$. While thse experiments at high impact velocities are relevant for meteorite and inter-asteroid impacts, the velocity ranges are not applicable to the landing or sample collection on a surface in low gravity.

Theoretical and experimental studies that are more relevant to an asteroid landing situation are found in the study of impactor deceleration in a granular medium: the \cite{katsuragi2007, katsuragi2013} and \cite{katsuragi2017} series studies the drag force exerted by a granular material on a penetrating impactor; \cite{clark2012} and \cite{clark2013} study the impactor stopping time and the energy transfer from the impactor to the granular material at the grain level in quasi-2D setups; \cite{seguin2008} studies the influence of container walls on the impactor penetration depth; and \cite{machii2013} measures the influence of the target strength on the impact outcome. All these experiments were performed at impact speeds ranging from a few 10 cm/s to a few m/s, and provide therefore relevant data for the exploration of asteroid surfaces. However, their experiments were performed in 1$g$, and the application of their findings to small body surfaces, which can have gravity levels $< 10^{-5}g$, remains unexplored.

The study of low-velocity impacts into granular material in reduced gravity was started in 1998 with the Shuttle experiment COLLisions Into Dust Experiment (COLLIDE) \citep{colwell_taylor1999Icarus, colwell2003Icarus}. Following experiments were performed during parabolic flight campaigns in 2002 and 2003, using the Physics of Regolith Impacts in Microgravity Experiment (PRIME) hardware setup \citep{colwell2008Icarus}. More recent data collection was obtained with these two experiments during the PRIME-3 \citep{colwell2016ASCE} and COLLIDE-3 campaigns, on the NASA C-9 aircraft in 2014 and Blue Origin's New Shepard suborbital rocket in 2016, respectively. Both these experiment setups generate impacts of $\sim$cm-sized spherical projectiles onto beds of granular material at speeds of 1-230 cm/s and gravity levels ranging from reduced gravity ($\sim10^{-2}g$) to microgravity ($<10^{-4}g$). The target materials used were quartz sand, JSC-1 Lunar and JSC Mars-1 simulants, sieved at particle size $< 250 \mu$m. While the results of each of these experiment campaigns has been published separately (except for the most recent COLLIDE-3 campaign), the collected data set has now reached a size of $\sim$130 data points and presents the opportunity for a new and combined analysis, with the perspective of applying these experimental results to interactions with the regolith-covered surfaces of small Solar System bodies.

This is the undertaking of the present paper. We combine the data sets of all the COLLIDE and PRIME campaigns and show the collision outcomes, coefficients of restitution, ejecta speeds, and estimated mass in dependance of the impact velocity and environmental $g$-level. Section~2 describes the PRIME and COLLIDE hardware setups and data collected. Section~3 describes the data analysis and results, which we then discuss in Section~4.

\section{Experimental setups}
\label{s:exp}

The data results presented here were collected over almost two decades during microgravity experimental campaigns on a variety of flight platforms. The hardware used to collect data can be divided into two families: (1) the COLLIDE experiment series, which flew on the Space Shuttle and a suborbital rocket, and (2) the PRIME experiment series, which flew on parabolic aircrafts.

\subsection{COLLIDE}

\begin{figure*}[tp]
  \begin{center}
  \includegraphics[width = 0.9\textwidth]{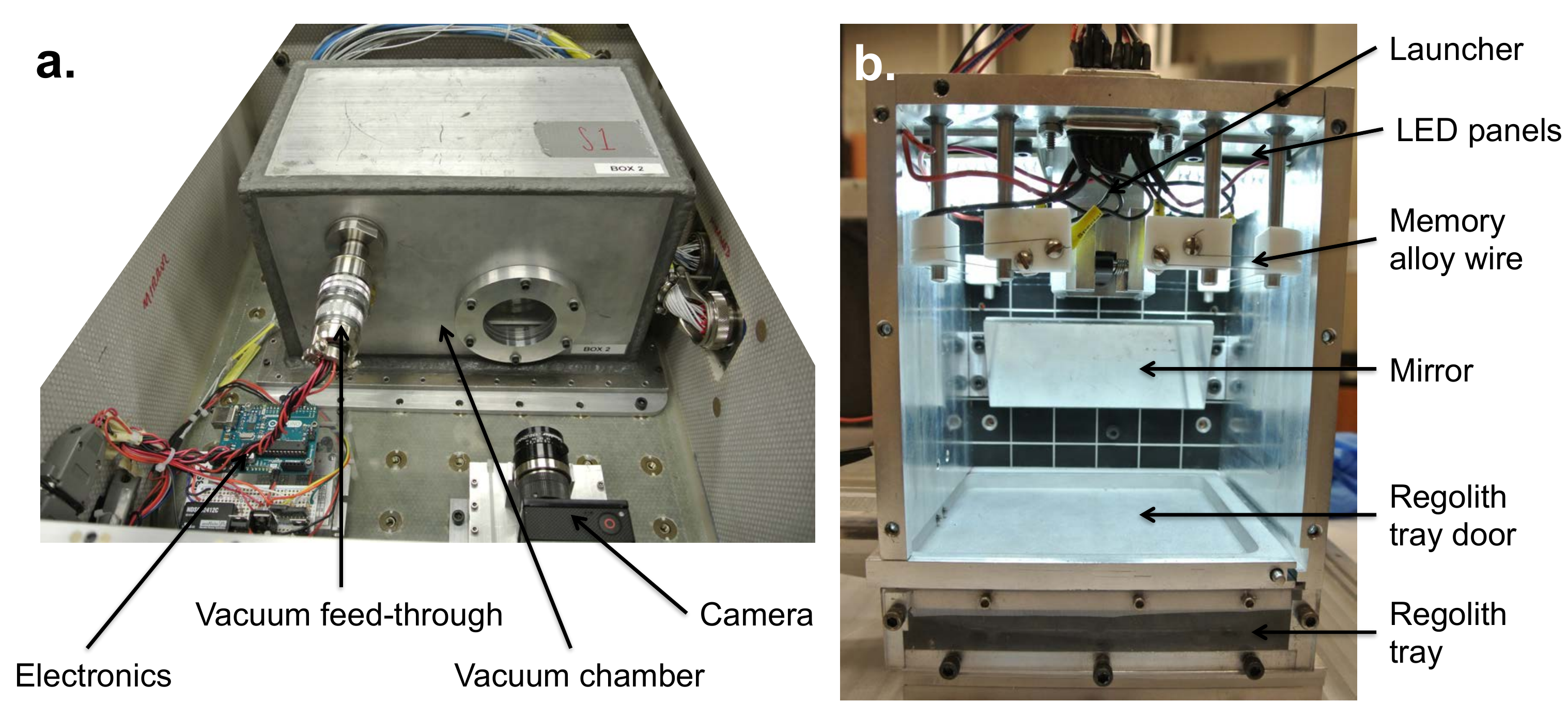}
 \caption{COLLIDE-3 hardware: (a) picture of the general setup showing the vacuum chamber with its viewport, the camera for data collecction, and the experiment electronics for autonomous experiment run; (b) picture of the Impactor Box System (IBS) located inside the vacuum chamber with the launcher and regolith tray.}
 \label{f:collide_hw}
 \end{center}
\end{figure*}

\paragraph{Flight campaigns}

The first two COLLIDE flight campaigns were performed in 1998 and 2001, on-board the Space Transportation System (Space Shuttle) STS-90 and STS-108 missions, respectively (COLLIDE-1 and -2). The experiment hardware and performance were described in \citet{colwell_taylor1999Icarus, colwell2003Icarus}. During these flights, a total of 9 impacts into regolith simulant were recorded.

More recently, in April 2016, a modified setup was flown on Blue Origin's New Shepard suborbital rocket (COLLIDE-3). Parts of the Space Shuttle COLLIDE hardware were repurposed for this new campaign, and a total of 4 impactor boxes were accommodated into the crew capsule of New Shepard. Three of these boxes were similar to the Space Shuttle hardware, launching one projectile into a bed of regolith simulant. The fourth box was outfitted with a multi-launcher instead of the single launcher configuration, now allowing for three impactors to be sequentially launched onto the bed of regolith simulant during the same experiment run. 

\begin{table*}[tbhp]
  \centering
  \caption{Projectiles and target materials of the three COLLIDE and the PRIME-3 campaigns. R = rebound with no ejecta; E = ejecta production, A = asteroid-level parabola (0.05$g$), X = no data available (regolith pillar). N/A indicates that no rebound from the projectile on the target could be observed. The table of the 108 PRIME-1 campaign impacts can be found in \cite{colwell2008Icarus}. No coefficients of restitution could be measured for the PRIME-1 impacts.}
    \begin{tabular}{|c|c|c|c|c|c|c|c|c|}
    \hline
      &  & \multicolumn{4}{c|}{Projectile} &  & & Coefficient\\ \cline{3-6}
        COLLIDE   &  & Diameter & Mass  & Material & Speed &Target & & of \\ 
        Campaign & IBS & [cm] & [g] & & [cm/s] & Material & Outcome & Restitution\\ \hline
    1     & 1     & 0.96  & 0.99  & Teflon & 14.80 & JSC-1 & R & 0.028\\
    1     & 4     & 0.96  & 0.99  & Teflon & 17.10 & JSC-1 & R & 0.022\\
    1     & 6     & 0.96  & 0.99  & Teflon & 90.00 & JSC-1 & R & 0.03\\
    2     & 1     & 2.00  & 10.71 & Quartz & 110.00 & Quartz Sand & E & 0.01\\
    2     & 2     & 2.00  & 9.64  & Quartz & 3.60  & Quartz Sand & E & N/A\\
    2     & 3     & 2.00  & 10.29 & Quartz & 1.29  & Quartz Sand & E & N/A\\
    2     & 4     & 2.00  & 8.66  & Quartz & 81.00 & Quartz Sand & E & 0.015\\
    2     & 5     & 2.00  & 8.98  & Quartz & 12.20 & JSC-1 & E & N/A\\
    2     & 6     & 2.00  & 10.62 & Quartz & 28.00 & Quartz Sand & E & 0.02\\
    3     & 2     & 2.05  & 10.55 & Quartz & 26.5 & Quartz Sand & R & 9.3$\times10^{-3}$\\
    3     & 3     & 2.05  & 10.55 & Quartz & 29.3 & JSC Mars-1 & E & N/A\\ \hline\hline
   PRIME-3 Flight  &  Box & &  &  &  & & & \\ \hline
    1     & 1     & 2.05  & 9.82  & Quartz & 48.45 & Quartz Sand  & E &  N/A \\
    1     & 3     & 2.05  & 11.76 & Quartz & 49.72 & Quartz Sand  & E &  N/A \\
    1     & 4     & 2.05  & 30.75 & Brass & 43.19 & Quartz Sand  & A &  N/A \\
    1     & 5     & 2.05  & 9.9   & Quartz & 43.75 & Quartz Sand  & E &  N/A \\
    1     & 6     & 2.05  & 28.2  & Brass & 19.87 & Quartz Sand  & X &  N/A \\
    1     & 7     & 2.05  & 9.82  & Quartz & 33.32 & JSC-1 & R & 0.078\\
    1     & 8     & 2.05  & 28.2  & Brass & 19.56 & JSC-1 & X &  N/A \\
    2     & 2     & 2.05  & 28.2  & Brass & 16.21 & Quartz Sand  & R & 0.046\\
    2     & 3     & 2.05  & 11.76 & Quartz & 31.52 & Quartz Sand  & A &  N/A \\
    2     & 4     & 2.05  & 30.75 & Brass & 38.4  & Quartz Sand  & A &  N/A \\
    2     & 5     & 2.05  & 9.9   & Quartz & 38.73 & Quartz Sand  & E & 0.025\\
    2     & 6     & 2.05  & 28.2  & Brass & 13.6  & Quartz Sand  & X &  N/A \\
    2     & 7     & 2.05  & 9.82  & Quartz & 5.55  & JSC-1 & R & 0.276\\
    2     & 8     & 2.05  & 28.2  & Brass & 14.45 & JSC-1 & R & 0.091\\
    3     & 1     & 2.05  & 9.82  & Quartz & 24.75 & Quartz Sand  & E &  N/A \\
    3     & 3     & 2.05  & 11.76 & Quartz & 26.38 & Quartz Sand  & E &  N/A \\
    3     & 4     & 2.05  & 30.75 & Brass & 31.84 & Quartz Sand  & A &  N/A \\
    3     & 5     & 2.05  & 9.9   & Quartz & 48.18 & Quartz Sand  & A &  N/A \\
    3     & 6     & 2.05  & 28.2  & Brass & 15.2  & Quartz Sand  & X &  N/A \\
    3     & 8     & 2.05  & 28.2  & Brass & 4.02  & JSC-1 & R & 0.229\\
    4     & 1     & 2.05  & 9.82  & Quartz & 25.85 & JSC-1 & E & 0.075\\
    4     & 3     & 2.05  & 11.76 & Quartz & 20.94 & Quartz Sand  & E & 0.020\\
    4     & 4     & 2.05  & 30.75 & Brass & 27.77 & JSC-1   & A &  N/A \\
    4     & 5     & 2.05  & 9.9   & Quartz & 52.93 & Quartz Sand  & A &  N/A \\
    4     & 6     & 2.05  & 28.2  & Brass & 10.02 & Quartz Sand  & X &  N/A \\
    4     & 8     & 2.05  & 28.2  & Brass & 13.38 & JSC-1   & R & 0.055\\ \hline
    \end{tabular}%
  \label{t:impact_list}%
\end{table*}%

\paragraph{Hardware}

\begin{figure}[tp]
  \begin{center}
  \includegraphics[width = 0.5\textwidth]{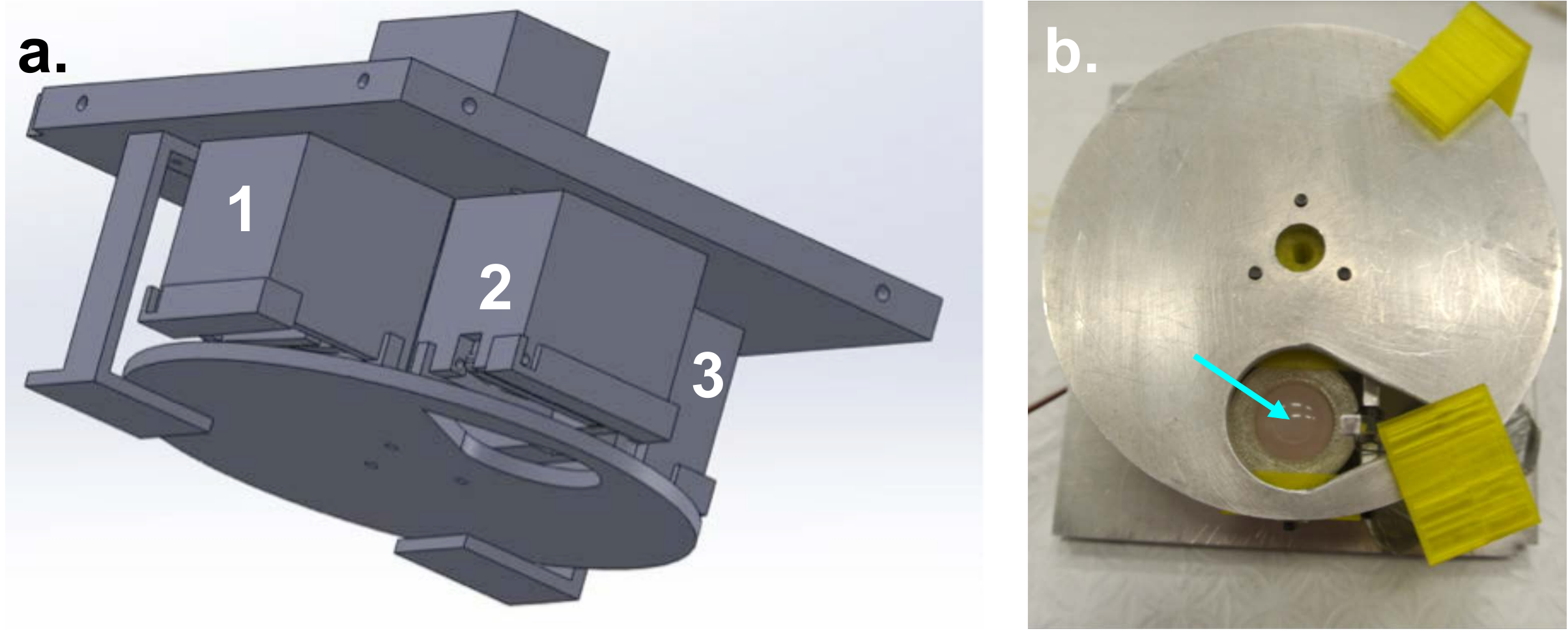}
 \caption{COLLIDE-3 multi-launcher system: (a) CAD side view showing the individual launchers (three, numbered) and the rotating plate at the bottom; (b) picture of the multilauncher from the bottom showing the rotating plate and a projectile (marked by an arrow) in its individual launcher through the plate's drop hole.}
 \label{f:collide_ml}
 \end{center}
\end{figure}

For COLLIDE-3, four individual Impactor Box Systems (IBSs) of COLLIDE-2 were modified for flight on suborbital launch vehicles. We manufactured a vacuum chamber for each IBS in order to perform the impacts under vacuum conditions, because the New Shepard crew capsule stays at atmospheric pressure during the flight. Each vacuum chamber was oufitted with a viewport allowing for a camera outside of the chamber to record the impact. The IBS electronics were re-wired and connected to custom electronics outside the chamber via an electrical feed-through. Both the chamber and IBS can be seen in Figure~\ref{f:collide_hw}. 

The camera system was entirely re-designed for the COLLIDE-3 campaign. While COLLIDE-1 and -2 used camcorders recording on digital video tapes, the new design used a GoPro Hero 3+ black edition recording data files on a microSD card. Both configurations required the retrieval of the tapes/memory card after the experiment run.

Three of the IBSs were used as single-launchers (one projectile per experiment run, like in the COLLIDE-1 and -2 campaigns), while the fourth one was modified to allow for the launching of three projectiles during the same experiment run. This modified setup was named multi-launcher. In a single-launcher IBS, the projectile is kept on a compressed spring by an aluminum bar. This bar is held in its closed position by a pin attached to two wires of memory shape alloy. When activated, these wires retract the pin and the released aluminum bar swings in its open position via a rotary spring. More details on the single-launcher setup can be found in \citet{colwell_taylor1999Icarus, colwell2003Icarus}.

The multi-launcher is composed of three individual launchers (slightly smaller than a single-launcher), a central stepper motor, and a circular bottom plate. The stepper motor allows for the rotation of the bottom plate, which is outfitted with a drop hole. The projectiles inside individual launchers are kept on a compressed linear spring. They are held in place by an aluminum bar forced in its closed position by the presence of the bottom plate. This bar is held under tension by a rotary spring pressing it towards an open position. When the drop hole aligns with the launcher, the pin is freed thus allowing for the linear spring of the launcher to push the projectile outwards. Figure~\ref{f:collide_ml} shows the multilauncher setup.

\paragraph{Projectiles and Simulants}
\label{s:coll_sims}

Table~\ref{t:impact_list} details which projectiles and simulants were used for all three experiment campaigns. The projectiles used in the COLLIDE campaigns were Teflon and quartz spheres of diameters between 1 and 2~cm. These projectiles were chosen in order to vary the impact energies over the largest range of values possible while using the same hardware setup. For a similar projectile size, 1-cm projectiles had masses about one order of magnitude smaller than 2-cm projectiles.

\begin{figure*}[t]
  \begin{center}
  \includegraphics[width = 0.95\textwidth]{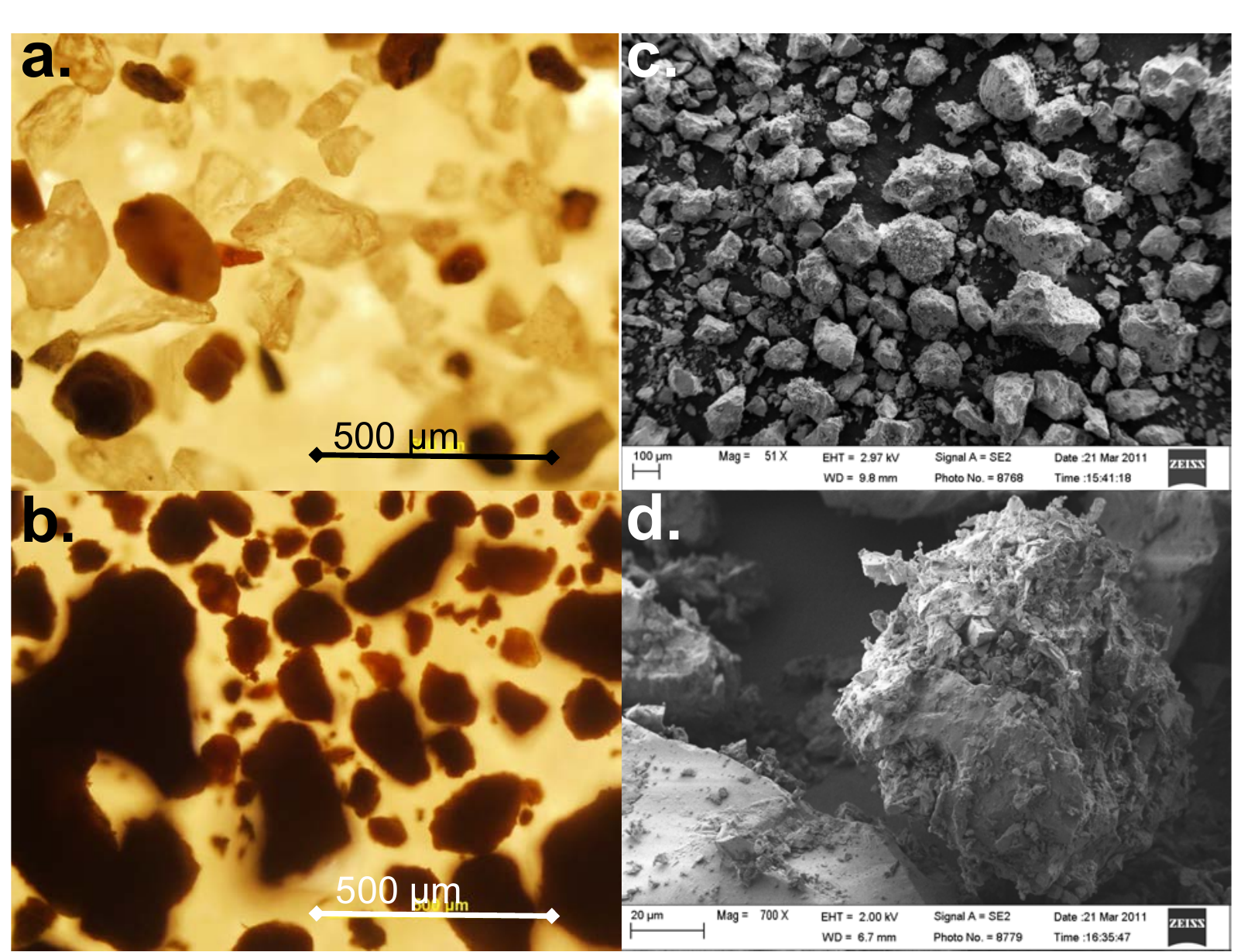}
 \caption{Optical and Sacnning Electron Microscope (SEM) pictures of the simulants used in COLLIDE and PRIME experiments: (a) quartz sand; (b) JSC Mars-1; (c) JSC-1 Lunar simulant. The sample were prepared in the same manner as for the experiments, by sieving the different grains to below 250 $\mu$m. (d) shows the details of the surface texture of JSC-1 grains.}
 \label{f:simulants}
 \end{center}
\end{figure*}

The target materials used were quartz sand, the lunar mare basalt simulant JSC-1, and the Mars soil simulant JSC Mars-1. The choise of these materials was not motivated by their mineralogic similarity to asteroid regolith, but rather by the shape of their grains. Indeed, from the relatively long history of inter-grain collision experiments \citep[see][for a review]{blum_wurm2008ARAA}, it appears that the actual mineral composition of the grains plays a comparatively negligible role in granular material cohesion compared to the size and shape of the grains. A notable exception to this observation takes place if the regolith is composed of ices (in particular water ice) or covered with organics, which are much stickier than rocky regolith and therefore change the behavior of grain layers. Ices and organic materials are outside the scope of the present work, and we therefore only considered grain shape and sizes. The main difference between the chosen simulants is the grain shapes and surface texture, which are increasingly angular and irregular from quartz sand to JSC Mars-1 and to JSC-1 Lunar simulant (\ref{f:simulants}). Note that quartz sand and JSC Mars-1 are more similar in shape compared with JSC-1.

The target materials were sieved to sizes between 75 and 250~$\mu$m, which provides for a normal distribution between these two sizes, and filled into the target tray of dimensions of 13 $\times$ 13 $\times$ 2 cm$^3$. Grain sizes on the surface of asteroids are known to have a size distribution ranging from $\mu$m to 10~m. The population of grains smaller than 1 cm is difficult to infer from available imaging of surfaces, such as for Vesta \citep{hiroi1994} and Eros \citep{dombard2010}, as the image resolutions are usually $>$ 1~cm. However, from thermal inertia measurements, \cite{gundlach_and_blum2013Icarus} has inferred average regolith grain sizes covering the surfces of 26 asteroids. For 15 of these 26 bodies, they determined an average grain size of $<$ 1 mm, which corresponds to the grain size distributions we used in the COLLIDE and PRIME experiments.

\cite{seguin2008} studied the influence of the grain container size on the penetration depth of projectiles droped into the container at speeds of $<$ 3 m/s. They showed that for a granular material target composed of 300-400 $\mu$m-sized grains and projectile sizes ranging from 5 to 40 mm, the container bottom had an influence on the projectile penetration depth only if located at a few grain diameters from the surface. They also showed that the container side wall location had a vanishing influence on the impactor penetration for a ratio between the tray and projectile of $>$ 5. Given their impact speeds and the size of their projectiles and target grains, their results are applicable to the COLLIDE and PRIME experiments. As the target trays were 13 $\times$ 13 cm$^2$ (13 $>$ 5*2 cm for a projectile of 2 cm in diameter) in lateral size, and 2 cm deep ($>>$ 250 $\mu$m grain diameter), we expect that the tray had no influence on the target response to the impacts performed in COLLIDE and PRIME.

The trays were consistently filled using the same procedure, in order to reproduce the same target density. This filling procedure consisted in the pouring of simulant material into the tray using a funnel, followed by the leveling of the surface using a flat metal ruler. We measured the porosities of targets produced in this manner in the laboartory (30 times) by filling the tray using this procedure, then measuring the mass of the target material using a scale. Knowing the container volume and the bulk density of the grains, we determined target porosities consistently ranging from 0.4 to 0.5. While the compaction level of fine grains at the surface of asteroids is not well determined, the bulk porosity of ``rubble-pile'' asteroids was estimated to be of this order by \cite{britt2002}, indicating that such porosity levels are relevant for the study of low-velocity impacts on asteroid surfaces. 

In the COLLIDE experiment setup, the target was contained by a sliding door, with no particular measures to prevent further compaction through post-preparation handling and launch vibrations. However, the COLLIDE-3 videos, which have a high enough resolution to show details of the target tray filling after the door opened during the rocket flight, show no volume reduction of the target sample. The target compaction during COLLIDE-1 and -2 could not be quantified due to the low resolution of the video data, but the projectile impact location indicate that the target surface was not significanlty shifted from its original position, so that we don't estimate porosities much below 0.4 during the experiment runs.

\subsection{PRIME}

\paragraph{Flight campaigns}

PRIME flew on three experiment campaigns on the NASA KC-135 and C-9 parabolic airplanes. The first flight campaign induced five flight weeks between July 2002 and May 2003 and is described in \citet{colwell2008Icarus}. The PRIME-1 campaign produced 108 successful impact recordings.

From the results of this first campaign it was clear that the regolith was very sensitive to the residual accelerations of the aircraft. Therefore, the PRIME-2 hardware was modified to be partially free-floating. In this setup, the individual impact boxes could be pulled out of the overall payload structure, which was bolted to the airplane floor. The boxes were still attached to the structure by an electrical umbilical. Unexpectedly, the residual accelerations of the plane during parabolas were still high enough to introduce torques onto the impact boxes via this umbilical, generating a worse acceleration environment than during PRIME-1 and no scientific data could be collected.

The PRIME-3 hardware was designed to be entirely free-floating. Each impact box had its own electronics attached to it, including a battery pack for power. The following paragraph describes the hardware flown during the PRIME-3 one-week flight campaign in August 2014. The PRIME-3 campaign produced 25 impact recordings. During the flight campaign, the pilots flew a number of parabolas at 0.05$g$ (named ``asteroid-level'' parabolas by the NASA flight support team) in addition to the usual microgravity ones. Out of the 25 recorded impacts, 7 were at asteroid level. Details on the PRIME-3 campaign can be found in \citet{colwell2016ASCE}.

\paragraph{Hardware}

\begin{figure}[tp]
  \begin{center}
  \includegraphics[width = 0.5\textwidth]{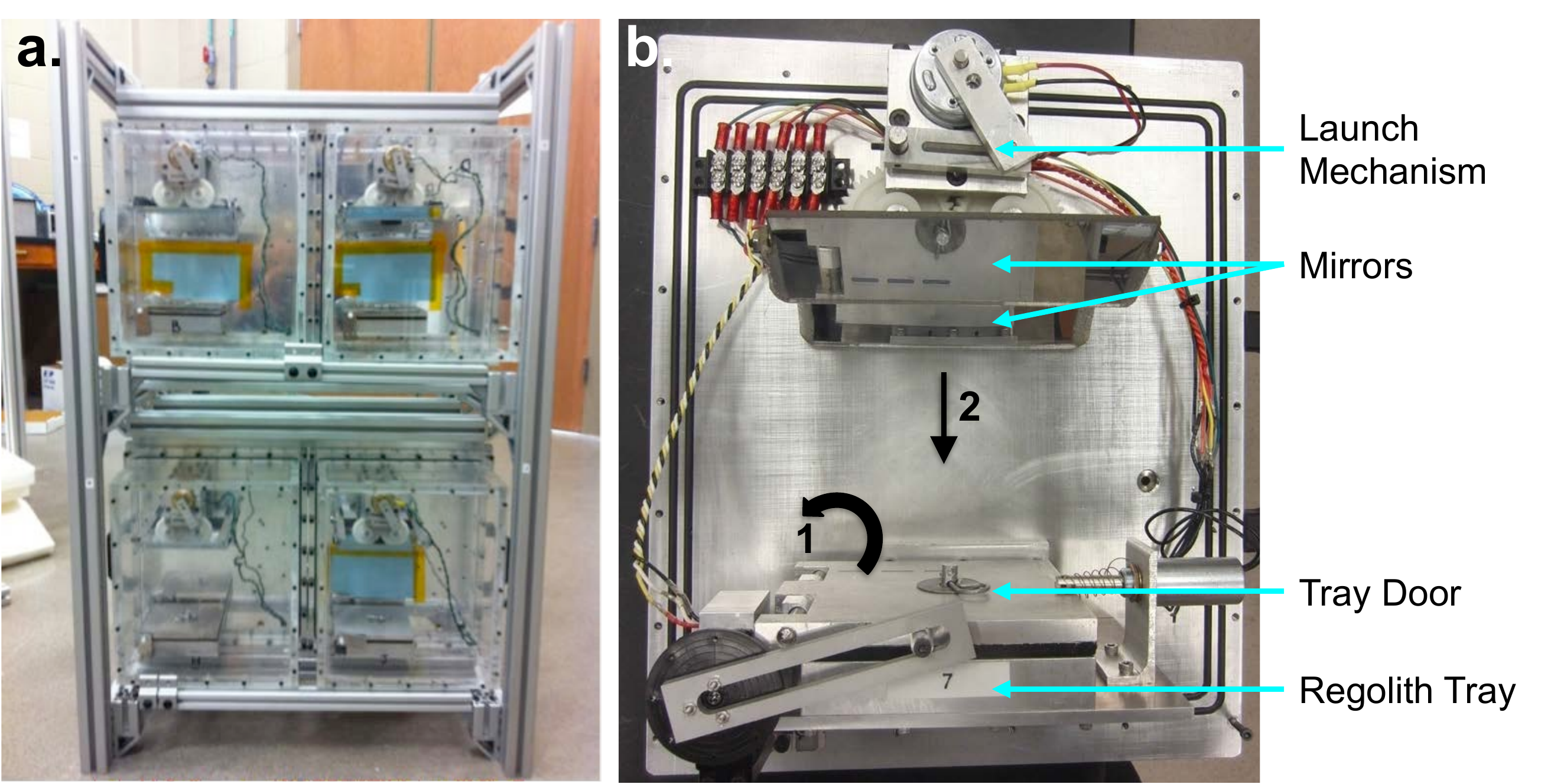}
 \caption{PRIME-3 hardware: (a) picture of the payload structure that is bolted to the aircraft floor and holds the 8 impact boxes (only 4 are visible in this front view); (b) interior of an impact box showing the regolith tray and the launch mechanism. During an experiment run, the tray door rotates to open (1) before the projectile gets launched towards the regolith bed (2).}
 \label{f:prime_hw}
 \end{center}
\end{figure}

Figure~\ref{f:prime_hw} shows the PRIME-3 hardware. It is composed of 8 individual impact boxes that are kept in a structure bolted to the airplane floor. Each box can be slid out of the structure and used to perform an experiment run while completely free-floating during a parabola. The impact boxes were inherited from PRIME-1 and -2 and modified to allow for an autonomous experiment performance: while the regolith tray and launch mechanisms are similar, the electronics were re-designed to actuate them externally using a wireless signal. One camera was used to record the impacts in the individual chambers as it was attached to a bracket that could be switched from box to box. The camera used was also a GoPro Hero 3+ black edition. More details on the inside of the impact boxes can be found in \citet{colwell2008Icarus}.

During asteroid-level parabolas, the impactor box was placed on the floor of the airplane and the projectile launch initiated at the beginning of the parabola.

\paragraph{Projectiles and Simulants}

The list of projectiles and simulants for the 108 impacts recorded during the PRIME-1 campaign can be found in \citet{colwell2008Icarus}. Table~\ref{t:impact_list} summarizes the projectiles and simulants used during the PRIME-3 campaign. 

The target materials and preparation methods were the same as for COLLIDE (see \ref{s:coll_sims} for details). The only difference with the COLLIDE targets is an additional contraption on the tray door that slightly compresses the target after preparation and through the pre-flight handling process. As the PRIME tray doors open by rotation rather than sliding, a plate was attached on the inner side facing the sample. Using four weak spring, this plate applies pressure on the target sample when the tray door is closed. In this way, the sample is jammed and its porosity remains unchanged (0.4 to 0.5 as described in \ref{s:coll_sims}) during loading of the experiment boxes on to the aircraft.

\section{Data analysis and results}
\label{s:results}

As the target size distribution and porosity were the same for all the COLLIDE and PRIME campaigns, we combined the collected data sets. This section presents how we analyzed the impacts and the results we obtained, integrating all the data collected. 

The platforms used to collect data in microgravity induced residual accelerations on the experiment hardware attached to them. These accelerations were around 10$^{-3}g$ in the Space Shuttle \citep{colwell_taylor1999Icarus}, and 10$^{-2}g$ on the parabolic aircraft \citep{colwell2003Icarus} and on Blue Origin's New Shepard rocket \citep{wagner2016}. The free-floating experiment boxes of PRIME-3 offered the best microgravity quality: with only air drag acting on the box moving at very low speeds ($<$ 1 mm/s), no residual acceleration could be detected from our video data. This implies that these experiment runs were performed at residual accelerations under our detection threshold, 10$^{-4}g$. In addition to these microgravity experiment runs, the PRIME-3 flights also allowed us to perform 7 experiment runs at 0.05$g$.

\begin{figure}[t]
  \begin{center}
  \includegraphics[width = 0.5\textwidth]{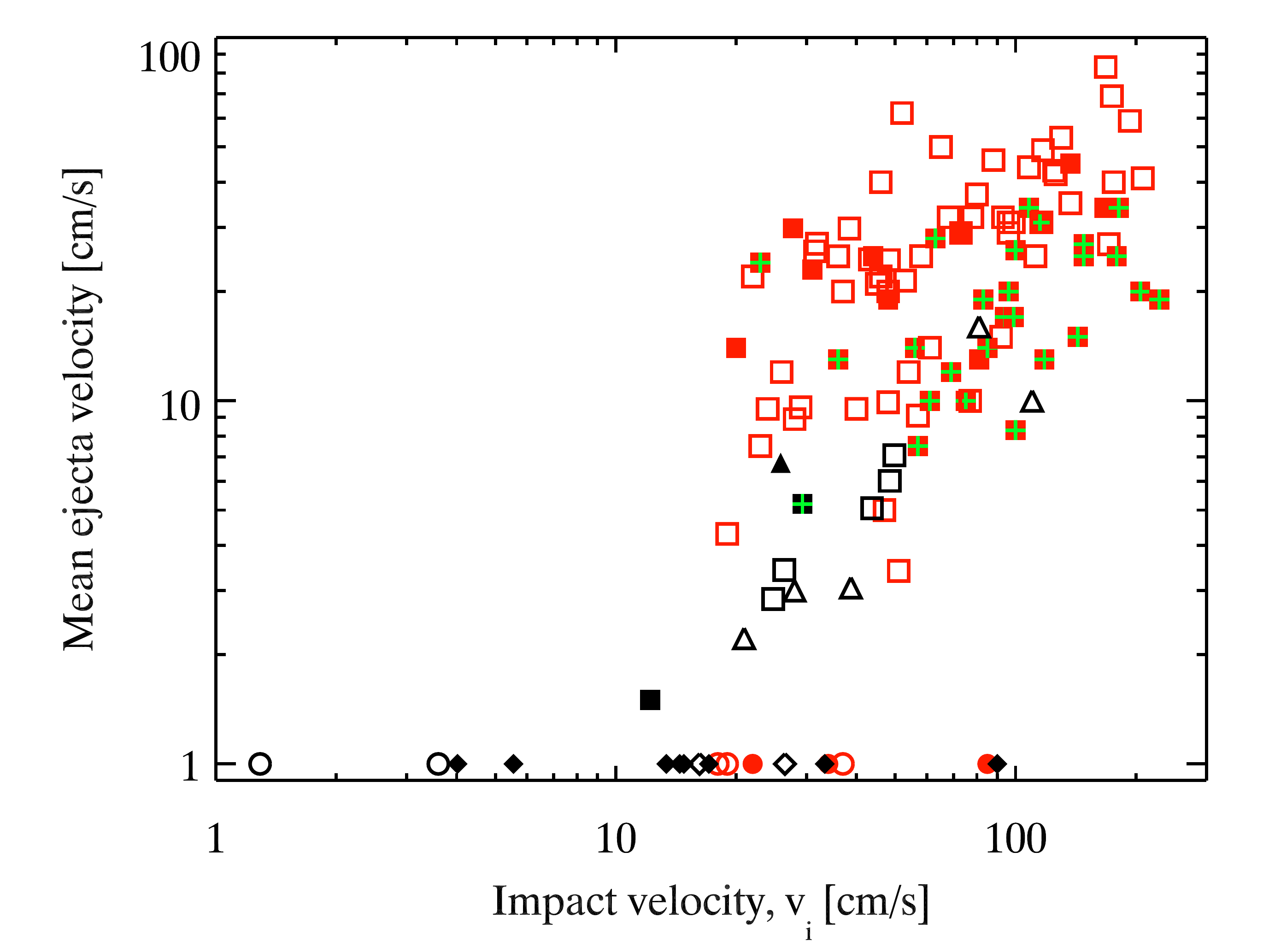}
 \caption{Outcomes of all the low-velocity impacts observed in PRIME-1, -3, and the COLLIDE campaigns as a function of the impact velocity. Circles: projectile embedded, no ejecta; diamonds: projectile rebound, no ejecta; squares: projectile embedded, ejecta production; triangles: projectile rebound, ejecta production. Open symbols: quartz sand target; filled symbols: JSC-1 target; green plus sign: JSC Mars-1 target. Red: reduced gravity ($\sim10^{-2}g$); black: microgravity ($<10^{-4}g$). All impacts producing no ejecta are shown at a mean ejecta velocity of 1 cm/s for the clarity of the graph, even though no speed was measured as no ejecta was observed.}
 \label{f:outcomes}
 \end{center}
\end{figure}

Across all the data collected, four types of collision outcome were observed:
\begin{itemize}
\item{Collisions producing ejecta with an embedded projectile (squares in Figure~\ref{f:outcomes}): in these collisions, no motion of the projectile after the impact is detected. Either the ejecta blanket or individual particles could be tracked and their speed determined.}
\item{Collisions producing ejecta with projectile rebound (triangles in Figure~\ref{f:outcomes}): the projectile can be tracked after the collision. In addition to the ejecta speeds, a coefficient of restitution could be measured for the projectile.}
\item{Collisions with no ejecta, but projectile rebound (diamonds in Figure~\ref{f:outcomes}). The coefficient of restitution for the projectile could be measured.}
\item{Collisions with no ejecta and embedded projectile (circles in Figure~\ref{f:outcomes}).}
\end{itemize}
Figure~\ref{f:outcomes} shows these outcomes as a function of the impact velocity. In this graph, all impacts not producing any ejecta were placed at a mean ejecta velocity of 1~cm/s. This is for the purpose of representing all the data on the same graph only, as no speed could be measured on absent ejecta (and a representation at 0 is not possible on a logarithmic scale). The following paragraphs give some additional details on these results for quartz sand and JSC-1 target materials. 

JSC Mars-1 behaved similarly to quartz sand in reduced gravity (the JSC Mars-1 data points seen in Figure~\ref{f:outcomes} are mostly from the PRIME-1 campaign, see Table 1 in \cite{colwell2008Icarus}). This can be explained by the similar grain shape between these two simulants (Figure~\ref{f:simulants} a. and b.). In microgravity, only one impact was available for data analysis, which was not enough for analyzing this set of data.

\subsection{Ejecta production}
\label{s:ejecta}

\begin{figure}[thp]
  \begin{center}
  \includegraphics[width = 0.5\textwidth]{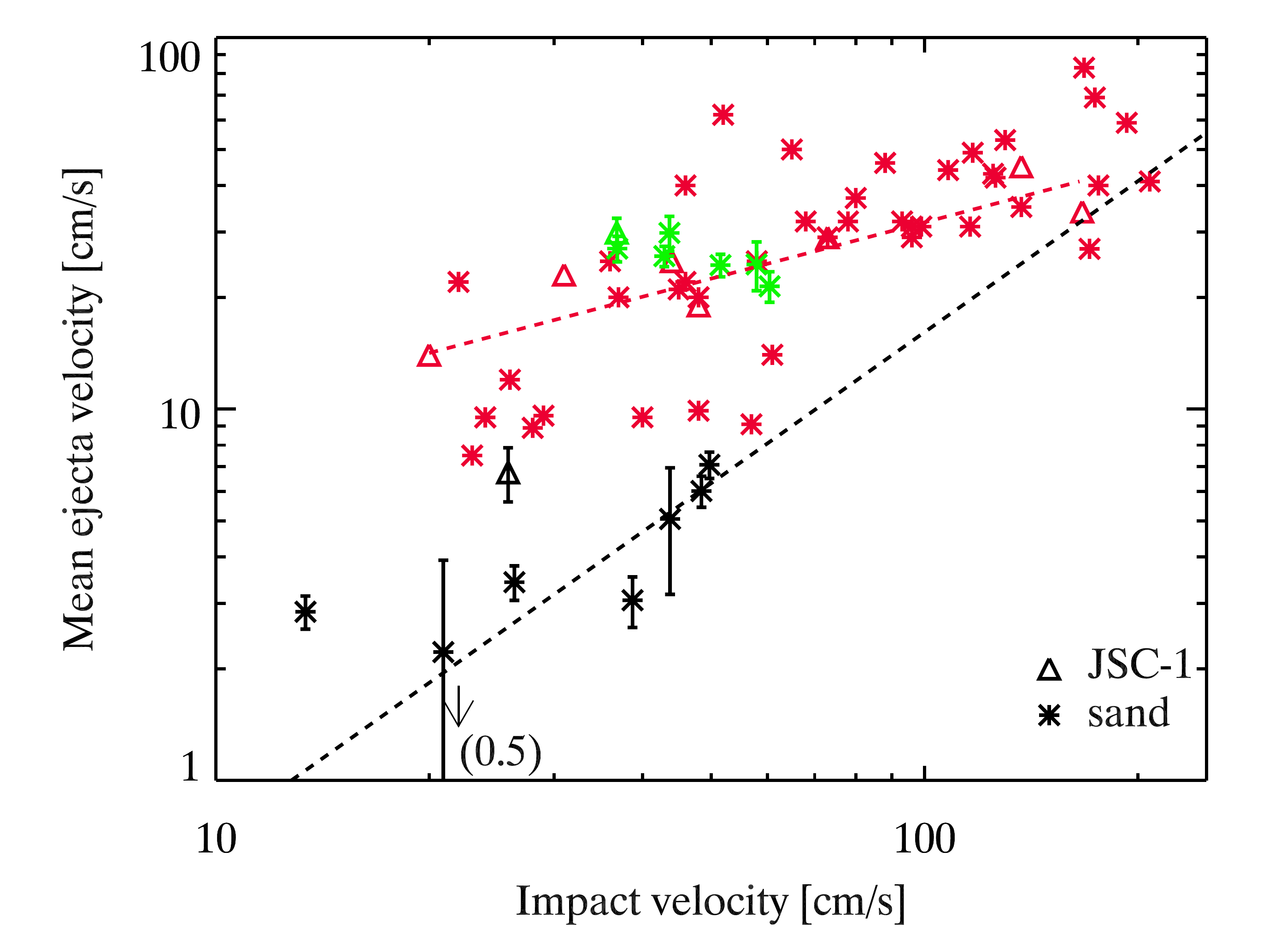}
 \caption{Combined PRIME-1and -3 ejecta velocity measurements as a function of the impact velocity. PRIME-1 data is shown by red symbols. For PRIME-3, black symbols show impacts in the free-floating box ($<10^{-4}g$), while green symbols show impacts at 0.05$g$. Impacts into quartz sand are marked by asterisks, and impacts into JSC-1 by triangles. Error bars are not shown for PRIME-1: they were smaller than the symbol size \citep{colwell2003Icarus}. PRIME-3 error bars represent the 1-$\sigma$ error of the Gaussian fit to the velocity distribution of tracked particles. The dashed black line is a power law fit to the microgravity impacts in sand (index 0.50), and the red dashed line is a power law fit to all the impacts into JSC-1 (index 1.35).}
 \label{f:ejecta}
 \end{center}
\end{figure}

\begin{figure}[t]
  \begin{center}
  \includegraphics[width = 0.49\textwidth]{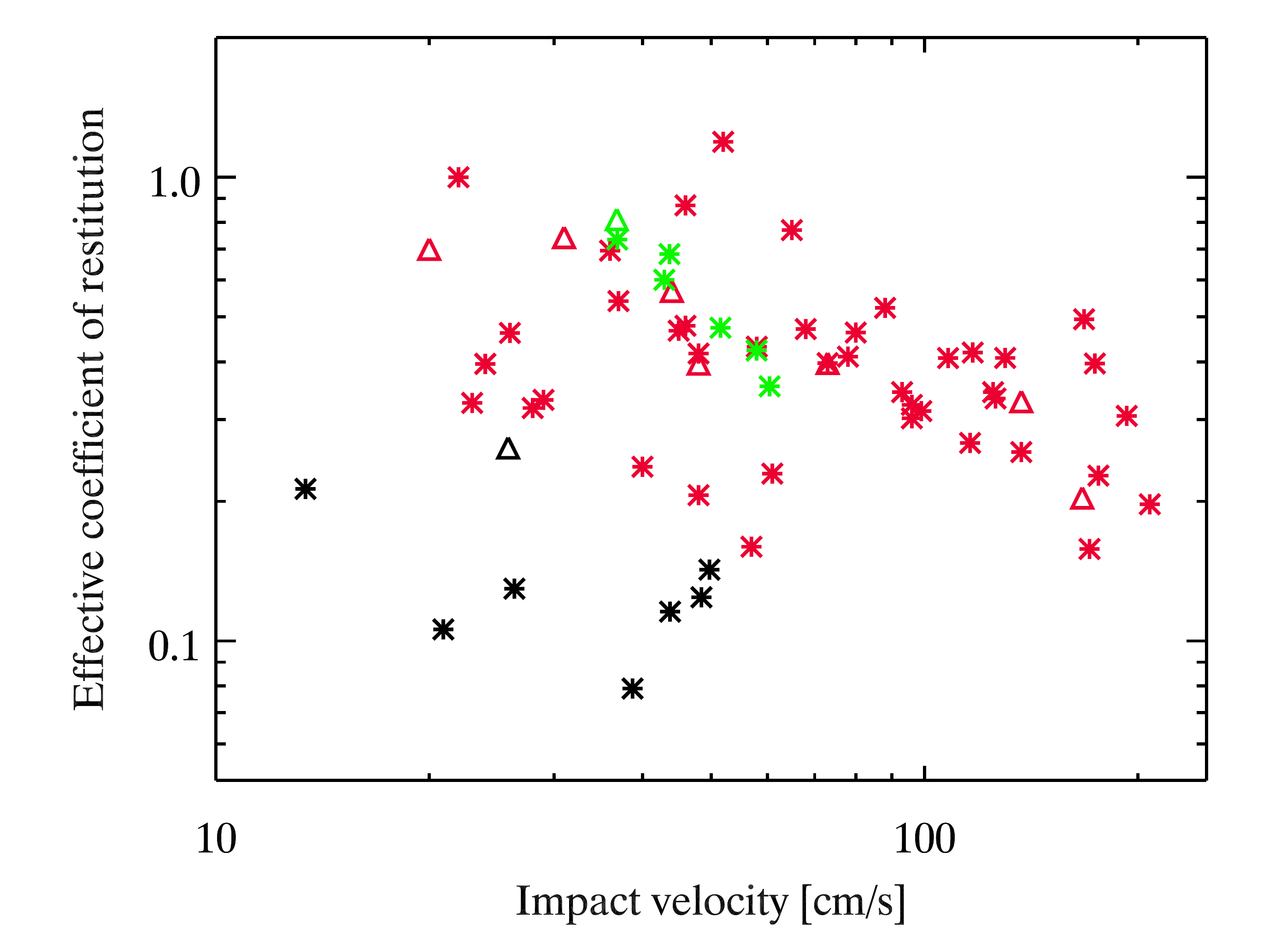}
 \caption{Effective coefficient of restitution of the impacts depicted in Figure~\ref{f:ejecta} (the same colors and symbols were used). This coefficient of restitution is defined as the ratio between the average ejecta velocity and the projectile impact velocity.}
 \label{f:eff_cor}
 \end{center}
\end{figure}

Impacts in COLLIDE-2, -3, PRIME-1, and -3 resulted in the production of an ejecta blanket. In the PRIME-1 and most of the COLLIDE-2 data \citep{colwell2003Icarus}, individual ejecta particles could not be distinguished due to the low camera resolution. However, features of the ejecta blanket, in particular the fastest particles forming the upper edge of the blanket (named ``corner'' in \citep{colwell2008Icarus}) were identified and allowed for the measurement of the upper end of the ejecta velocity distribution. 

In PRIME-3, the camera resolution allowed for the individual tracking of ejecta particles. For these experiments, we were able to measure a mean ejecta velocity by directly tracking 30 particles for each impact, using the Spotlight software developed at NASA Glenn Research Center \citep{klimek2006}. However, due to the nature of the video data collected, the tracking of individual particles is limited to optically thin portions of the ejecta blanket. This includes but is not restricted to the ejecta ``corner'' mentioned above. Particle tracking was performed by three people independently in order to reduce the measurement errors. The ejecta grain velocity distribution obtained for each impact were normal and could be fit by a Gaussian curve. Details on the tracking methods can be found in \cite{colwell2016ASCE} (see their figures 5 and 6). For PRIME-3 and COLLIDE-3, the characteristic ejecta velocity was chosen to be the Gaussian mean of velocity distribution of the tracked particles.

Figure~\ref{f:ejecta} shows the results of the ejecta speed measurements for PRIME-1 and -3. The 1-$\sigma$ error bars for the Gaussian mean determination are shown for the PRIME-3 data points. COLLIDE-3 produced ejecta for only one impact and the target material was JSC Mars-1. As stated above, JSC Mars-1 behaved similarly to quartz sand and we are not including JSC Mars-1 results here. The plateau formed by the PRIME-1 microgravity data points (black asterisks representing quartz sand impacts) around 10~cm/s for impact energies below 10$^4$~ergs is due to the residual accelerations induced by the parabolic aircraft on the experiment boxes (on the order of 10$^{-2}g$). Only particles ejected with energies able to overcome the effect of this ambient gravity field were lifted from the target. For this reason, the slowest moving ejecta has a minimum speed, set by this $g$-level, and the measured velocities follow a plateau. This plateau also formed in PRIME-3 mirogravity data, but at ejecta velocities around a few cm/s, indicating that residual accelerations were successfully reduced by the free-floating hardware configuration, compared to PRIME-1, which was attached to the aircraft frame.

The scatter of the PRIME-1 data points towards higher ejecta velocities was induced by the observational bias produced by the tracking method used: tracking the ejecta ``corner'' described above limits the recorded speeds to the fastest moving particles, and is therefore only representative of the upper end of the ejecta velocity distribution: there is therefore a tendency to measure higher ejecta velocities at the same impact energies. However, the data points show a trend followed by the lower limit of the measured velocities: to guide the eye, we fit a power law to the impacts into quartz sand in microgravity. The index obtained in 1.35. For all impacts into JSC-1, a fitted power law index is at 0.50. We collected only one data point for JSC-1 ejecta at microgravity levels, which lies below this power law fit. Further data collection on JSC-1 ejecta-producing impacts in microgravity will be required in order to evaluate if this point is indicative of a different ejecta behavior between the two gravity levels.

The ejecta speeds measured for impacts at 0.05$g$ during the PRIME-3 campaign are of the same range as the ones measured during the PRIME-1 campaign, where microgravity parabolas were flown, but the experiment hardware was fixed to the aircraft frame. This demonstrates that the residual accelerations produced on the experiment when attached to the aircraft are comparable to 10$^{-2}g$, and that our method for determining the average ejecta speeds provide similar values than the method used in \cite{colwell2008Icarus}.

We defined the effective coefficient of restitution of the impacts as the ratio between the average ejecta velocity and the projectile impact velocity, and investigated its relationship to the impact velocity. These effective coefficients of restitution showed no correlation and a uniform distribution over the available range of impact velocities (Figure~\ref{f:eff_cor}). We calculated the overall average effective  coefficient of restitutionto be 0.39 $\pm$ 0.15. This value is about twice the one derived by \cite{deboeuf2009} for impacts of cm-sized spheres into 100 $\mu$m glass bead beds. Deviations to our mean value were much higher than for the measurements performed in \cite{deboeuf2009}. When separating the impacts into quartz sand from the ones into JSC-1, we obtained average effective coefficients of restitution of 0.38 $\pm$ 0.15 and 0.45 $\pm$ 0.16, respectively. When separating between impacts in low- and microgravity, we obtained values of 0.43 $\pm$ 0.14 and 0.15 $\pm$ 0.04, respectively.

\begin{figure}[t]
  \begin{center}
  \includegraphics[width = 0.5\textwidth]{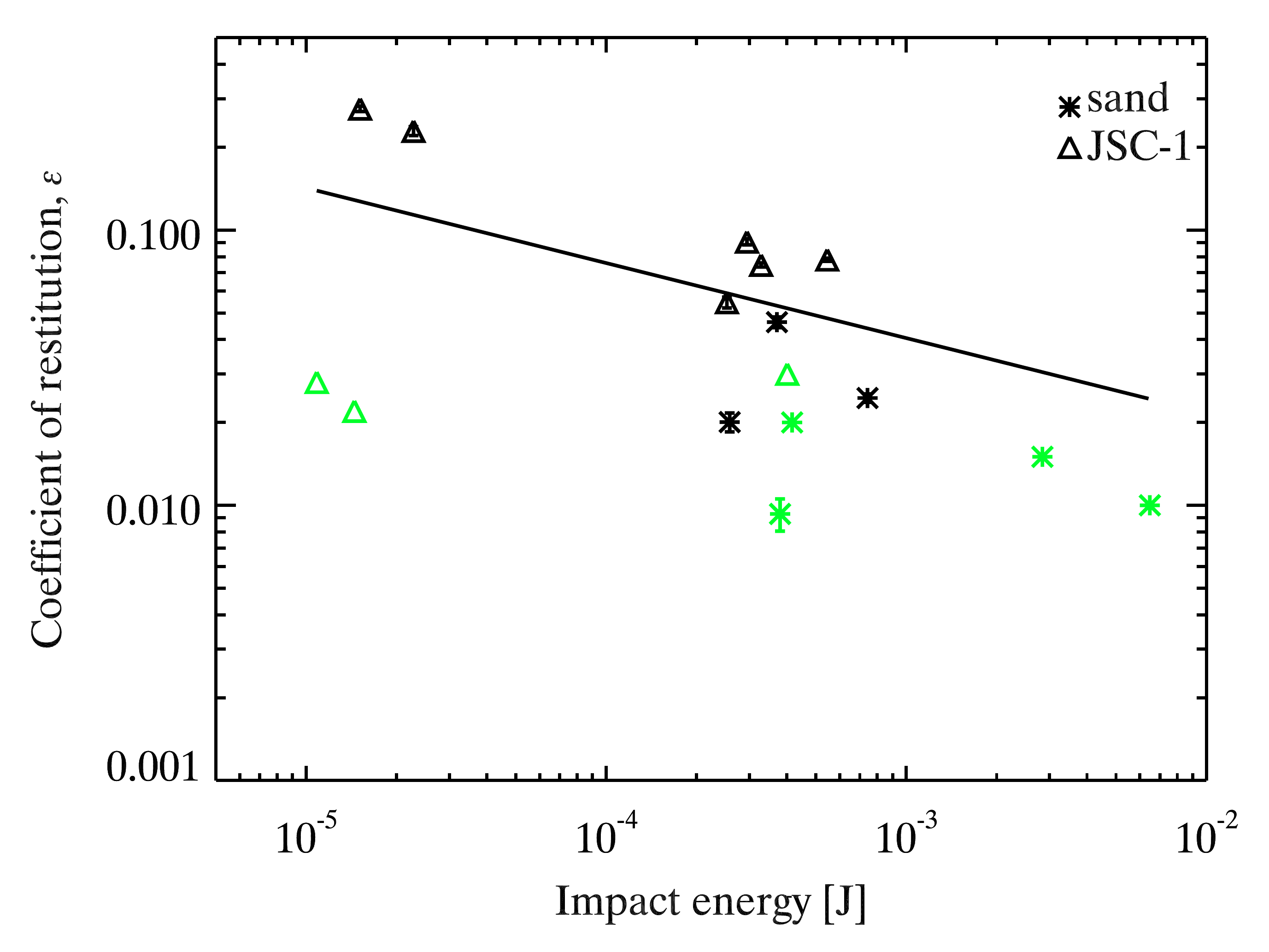}
 \caption{Combined PRIME-3 (black) and COLLIDE (green) coefficient of restitution measurements. This coefficient of restitution is defined as the ratio of the projectile speed after and before the impact with the regolith bed. Impacts into quartz sand are marked by asterisks, and impacts into JSC-1 by triangles. Error bars are shown to for the PRIME-3 and COLLIDE-3 measurements. For COLLIDE-1 and -2 they can be found in \citet{colwell2003Icarus}. The straight line represents the power law fit to all data points shown, which has an index of -0.27.}
 \label{f:COR}
 \end{center}
\end{figure}

\subsection{Projectile rebound}
\label{s:rebound}

A number of impacts during the COLLIDE-1, -3, and PRIME-3 campaigns showed a rebound of the projectile after impact on the target and allowed for the measurement of a coefficient of restition $\epsilon=\frac{v_f}{v_i}$, $v_i$ and $v_f$ being the projectile velocities before (initial) and after (final) the impact (see Table~\ref{t:impact_list}). Figure~\ref{f:COR} shows the measured coefficients of restitution as a function of the impact energy. For speeds above 30~cm/s, impacts systematically produced an ejecta blanket and coefficients of restitution of the projectile could either not be measured or were of the order of 10$^{-2}$.

At impact speeds between about 20 and 30~cm/s, both ejecta production and projectile rebound without ejecta were observed. Below 20~cm/s, only rebounds without ejecta were observed. In order to compare our results to the experimental work of \cite{katsuragi2017}, we fit a power law to this set of data.In  \cite{katsuragi2017}, spherical projectiles of about 1 cm in diameter are dropped into aggregates composed of 750 $\mu$m SiO$_2$ monomer grains with a packing density of $\Phi$=0.35. The index we obtain with our data is -0.27, similar to the value of -1/4 found by \cite{katsuragi2017}. We note, however, that if we separate the data sets from COLLIDE and PRIME, we obtain indexes of -0.10 and -0.50, respectively, a difference mostly due to the two very low energy impacts into JSC-1 of the COLLIDE-1 flight. As mentioned in \cite{katsuragi2017}, this overall index value is twice as high as expected from the theory of an elastic sphere impacting a plane surface \citep{johnson1985,thornton1998}. This indicates that energy absorption in a bed of granular material is not entirely captured by the mechanics of elastic surfaces.

\begin{figure}[t]
  \begin{center}
  \includegraphics[width = 0.5\textwidth]{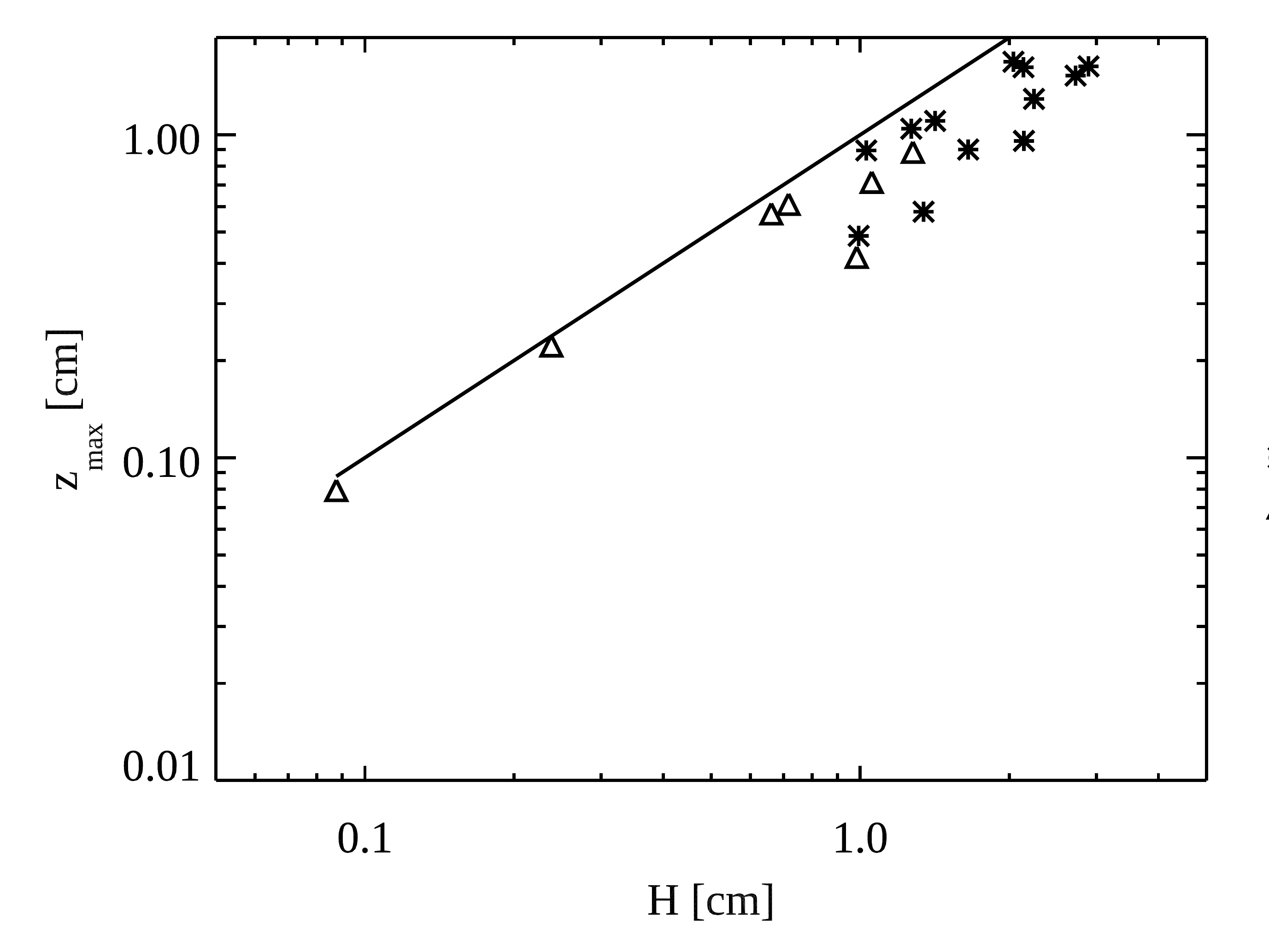}
 \caption{Maximum penetration depth of the projetile into the target for PRIME-3 impacts. H is the equivalent total drop distance as described in the text. Impacts into sand are marked by asterrisks, and impacts into JSC-1 by triangles. The line represents an index 1 power law.}
 \label{f:pd}
 \end{center}
\end{figure}

\subsection{Projectile penetration depth}
\label{s:pd}

In the PRIME-3 video data, the resolution is high enough to determine the maximum penetration depth of the projectile into the target for 21 impacts. This maximum penetration depth is defined as the distance between the bottom of the projectile and the target surface. Figure~\ref{f:pd} shows these results as a function of the equivalent total drop distance $H$. This distance was derived to be able to compare our results with the ones obtained by \cite{katsuragi2017}: $H = h + z_{max}$, where h is the drop height (experiments in \cite{katsuragi2017} are performed in 1$g$, and the drop height is used to control the impact velocity), and $z_{max}$ is the maximum penetration depth of the projectile into the target. As the impact velocity was not determined by the drop height in PRIME experiments, but by the stored potential energy in the spring of the projectile launcher, we calculate an equivalent $h$ from the impact velocity $v_i$ by equating the 1$g$ potential energy with our kinetic impact energy as follows:

\begin{equation}
\begin{aligned}
E_{1g} = E_{0g} \\
mgh = \frac{1}{2}mv_i^2 \\
h = \frac{v_i^2}{2g}
\end{aligned}
\end{equation}

with $g$ = 9.81 m/s. As shown by the index 1 power law in Figure~\ref{f:pd}, $H$ is dominated by $z_{max}$ in half of the impacts. This is due to the very low impact velocities, inducing equivalent drop heights $h \ll z_{max}$, so that $H \sim z_{max}$. We do not observe any correlation between the maximum penetration depth and the computed total equivalent drop distance. This is in contrast with the relation $z_{max} \propto H^{1/3}$ observed in \cite{katsuragi2017}, representative of a scaling by impact energy. A scaling by momentum of the form $z_{max} = Am_p^{\alpha}v_i^{\beta}$, with $m_p$ and $v_i$ being the projectile mass and impact velocity, respectively, and $A$, $\alpha$, and $\beta$ fit parameters, as performed in \cite{guettler_et_al2009ApJ}, was not successful either.

\begin{figure}[t]
  \begin{center}
  \includegraphics[width = 0.5\textwidth]{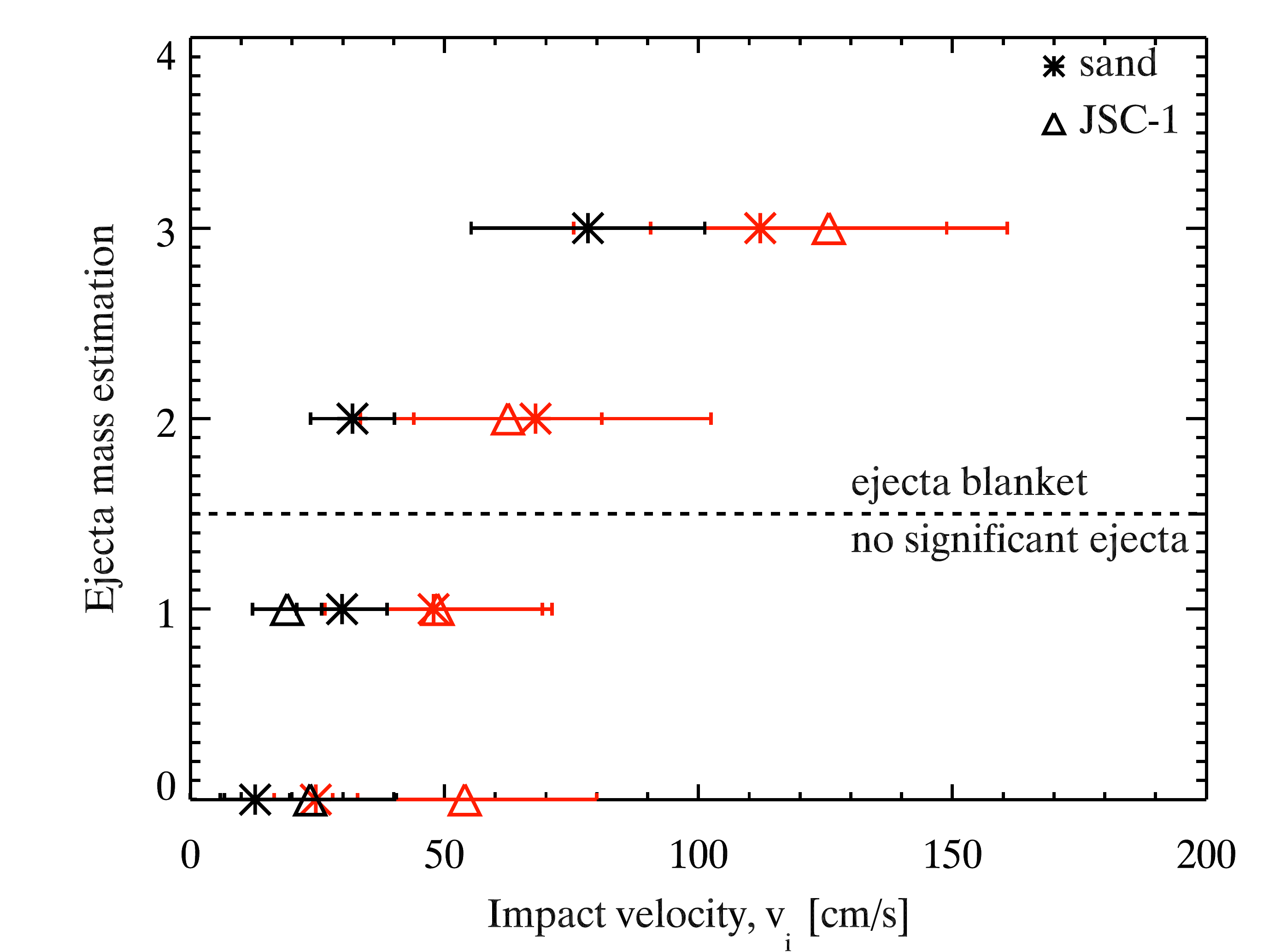}
 \caption{Arbitrary quantified amount of ejecta for all collected data on sand and JSC-1 targets. The amount of ejecta mass was quantified as follows: 0 = no ejecta; 1 = ejection of single particles; 2 = ejecta blanket, mass of the order of the projectile mass; 3 = ejecta blanket, mass much higher than the projectile mass. The impact velocity given for each target type is the average of all impacts that created the same amount of ejecta. Asterisk = quartz sand target; Triangle = JSC-1 lunar simualnt target. Black = microgravity impacts; Red = reduced-gravity impacts. The dotted line marks the limit between no significant ejecta produced (either none at all or only individual particles detaching) and an ejecta blanket.}
 \label{f:mass}
 \end{center}
\end{figure}

\subsection{Grain behavior with decreasing gravity level}

From the experimental results presented here, we observed differences in the response of the target to low-velocity impacts in reduced gravity ($\sim10^{-2}g$) and microgravity ($<10^{-4}g$). Figure~\ref{f:outcomes} shows that compared to the total number of impacts in reduced gravity ($\sim10^{-2}g$), only a few of them did not produce ejecta: the vast majority of impacts generated an ejecta curtain. In addition, none of the reduced gravity impacts lead to a rebound of the projectile (no red triangles or diamonds). In particular, when no ejecta was produced around the lowest impact speeds oberved ($\sim$20 cm/s), only projectile embedding in the target could be seen. For the same impact speeds, a microgravity environment ($<10^{-4}g$) lead to the rebound of the projectile. In fact, embedding into the target was only seen in the two impacts under 4 cm/s. All other impacts than those without ejecta showed a rebound of the projectile.

Projectile rebound was also observed in combination with an ejecta curtain in half of the impacts observed in microgavity (black triangles). Even though Figure~\ref{f:COR} shows that the coefficient of restitution of these rebounds is much smaller compared to the ones with no ejecta production (diamonds), this behavior demonstrates the differences in target response in reduced- and microgravity environments.

It can also be noted that for the same impact speed, ejecta is faster on average in reduced gravity compared to microgravity (red symbols are above black symbols in Figure~\ref{f:outcomes}). In addition, less mass is ejected than in microgravity. Figure~\ref{f:mass} shows the average ejecta speeds for four levels of ejected mass: as the ejecta mass can not be measured after each experiment run (the return to 1$g$ mixes the target material from the tray and the ejected particles), an estimation was performed from the video images. Four levels of ejecta mass were identified: no ejecta (0); individual particles detaching from the target (1); ejected mass of the order of the projectile mass (2); ejected mass much higher than the projectile mass (3) (we followed the same numbering as in \cite{colwell2003Icarus}). In Figure~\ref{f:mass}, the red symbols show reduced-gravity impacts ($\sim10^{-2}g$), and black symbols microgravity impacts ($<10^{-4}g$). The impact velocity given for each target type and ejecta mass level is the average of all impacts that created the same amount of ejecta. From the sand targets, we can see that the same average impact speed results in higher ejecta masses in microgravity compared to reduced gravity. Together with Figure~\ref{f:outcomes}, this shows that fewer target particles are ejected in reduced-gravity, but their ejection speed is higher.

Figure~\ref{f:mass} also shows that no ejecta blanket was observed in microgravity with JSC-1 targets. While this is also due to the fact that only very few impacts were performed in JSC-1 in microgravity at speeds $>$50 cm/s, the nature of the target plays a role in the ejecta mass produced: compared to quartz sand particles, which are rounded and considered cohesionless in vacuum (once the air humidity is removed), JSC-1 particles are more angular and behave like a cohesive powder. Figure~\ref{f:outcomes} shows that only two out of eight impacts into JSC-1 at $>$10 cm/s resulted in ejecta production, and these were only individual particles detaching instead of an ejecta blanket. As we can see in Figure~\ref{f:mass}, the impact speed does not seem to influence the production if these individual particles in JSC-1 targets, compared to quartz sand targets, which display a more consistent trend of increasing ejecta mass with increasing impact velocity. When an ejecta blanket is produced though, we can see that higher impact velocities result in higher ejected masses for both quartz sand and JSC-1 targets.

\section{Discussion}
\label{s:discussion}

\subsection{Observing ejecta}

\begin{figure}[t]
  \begin{center}
  \includegraphics[width = 0.47\textwidth]{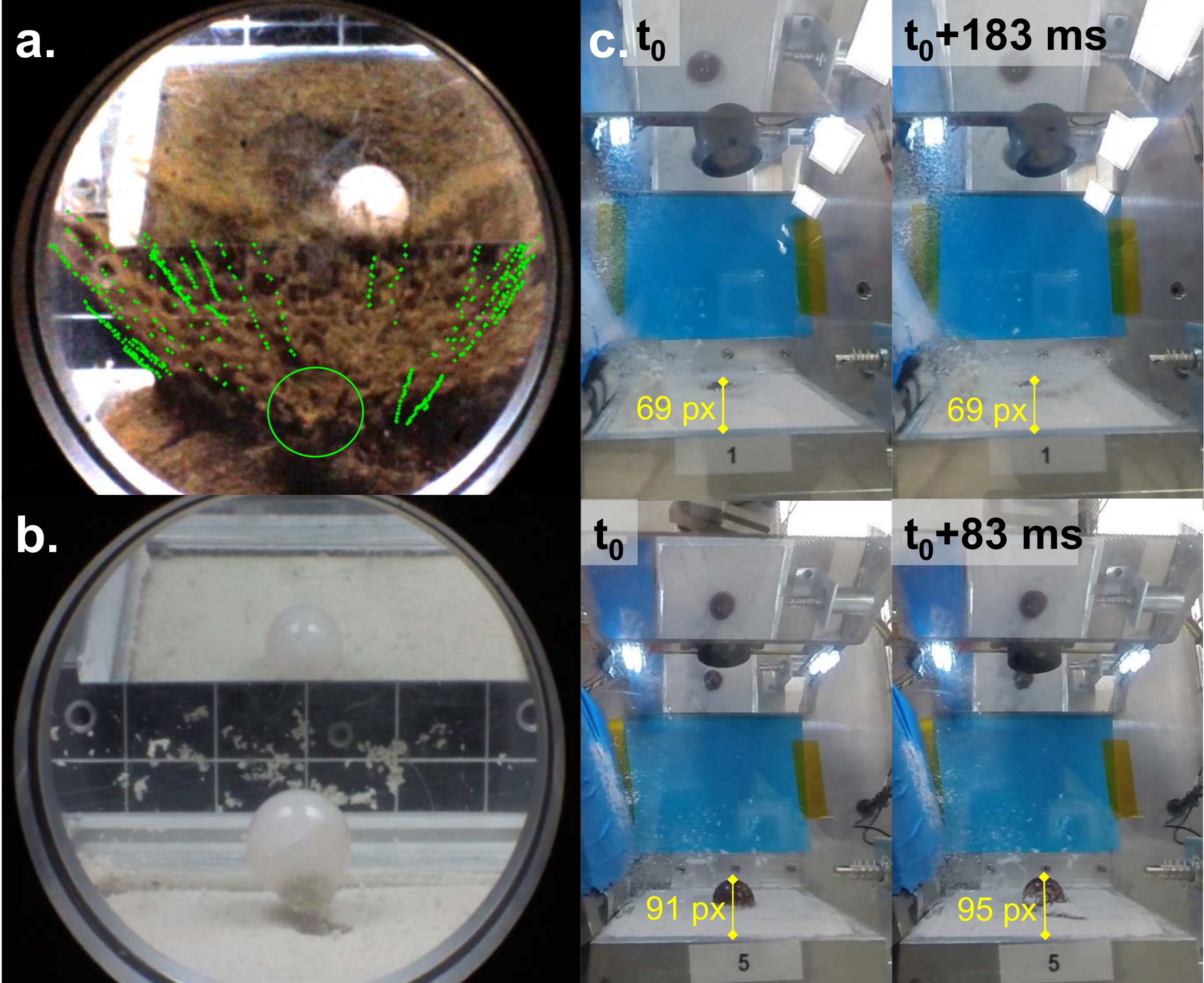}
 \caption{Data analysis examples: (a) COLLIDE-3 IBS 3 impact into JSC Mars-1 at 29.3 cm/s. Particle tracks are superposed in green and the approximate projectile position is shown with a green circle, as the ejecta curtain hides it entirely. The mirror above the target tray shows a top view of the impact, in which the projectile is visible. (b) COLLIDE-3 IBS 2 impact into quartz sand at 26.3 cm/s. Note the mass transfer onto the projectile after rebound. (c) PRIME-3 rebound observation: t$_0$ indicates the moment of deepest penetration of the projectile into the target. (top) Case of ejecta production without projectile rebound (flight 1, box 1): no marble motion can be detected for 183 ms. In following frames, the ejecta curtain hides the projectile, which can not be tracked anymore. (bottom) Case of ejecta production with projectile rebound (flight 2, box 5): we can track the motion of the projectile over 10 frames and 4 pixels.}
 \label{f:analysis}
 \end{center}
\end{figure}

Due to the hyper-$g$ levels experienced after the low gravity phase, the state of the target and projectile after each impact is destroyed and the PRIME and COLLIDE data collection relies solely on video recordings. In particular, it is not possible to measure the ejected mass, nor the trajectories of all ejected particles in the case of an ejecta blanket forming. As described in \ref{s:ejecta}, ejected particles could only be tracked in thin parts of the ejecta blanket, or in the case of single particles being lifted rather than a blanket. Figure~\ref{f:analysis}a shows an example of particle tracks superposed with the ejecta blanket: even though we can get particle speeds from these partial tracks, there is a high uncertainty in placing the origin of each track with respect to the surface of the target material and embedded projectile, which location can only be estimated as it is obscured by the ejecta blanket. For this reason, it is not possible to generate mass vs. speed or speed vs. launch position diagrams for ejecta, in order to compare them to 1-$g$ data and the scaling laws derived in \cite{housen2011}.

\subsection{Observing projectile rebound}

\begin{figure*}[t]
  \begin{center}
  \includegraphics[width = 0.97\textwidth]{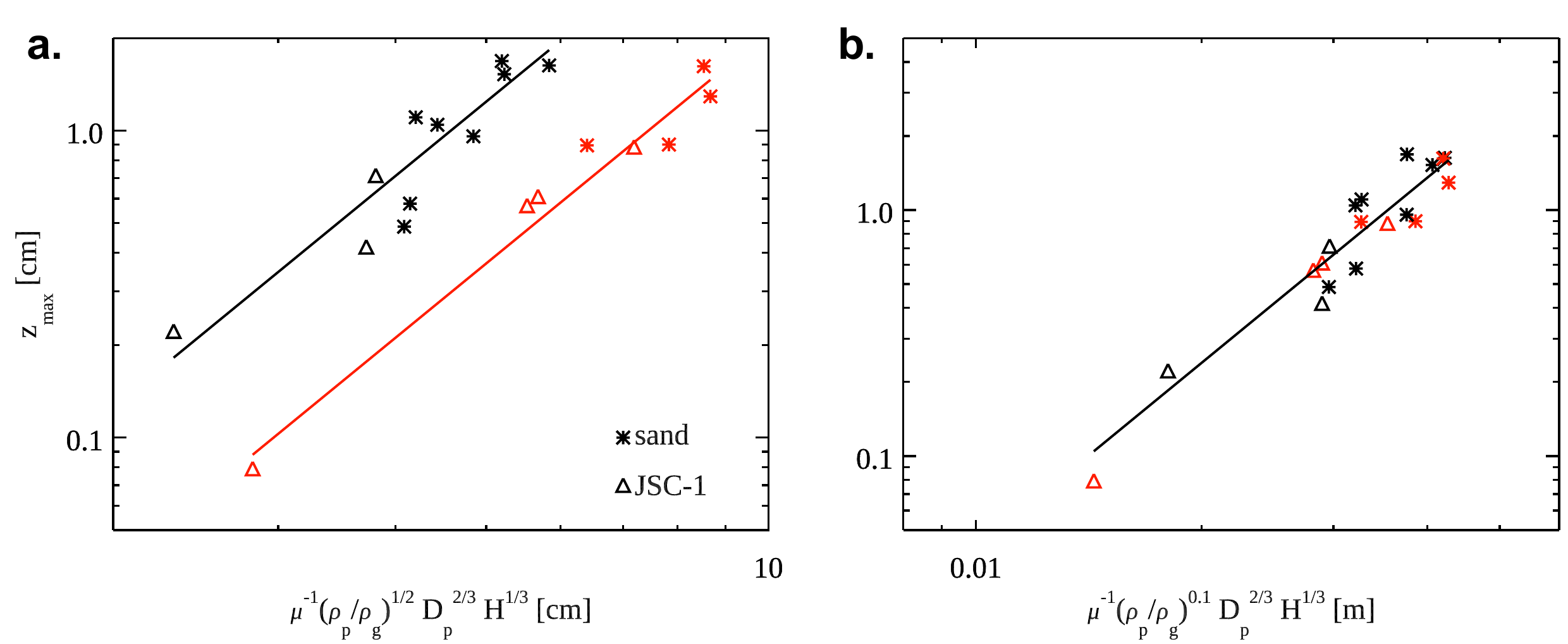}
 \caption{Scaling of the maximum penetration depth during PRIME-3 impacts: (a) with the quantity $\mu^{-1}(\rho_p/\rho_g)^{1/2}D_p^{2/3}H^{1/3}$ (see text for details); and (b) with $\mu^{-1}(\rho_p/\rho_g)^{0.1}D_p^{2/3}H^{1/3}$. Asterisks and triangles represent quartz sand and JSC-1 targets, respectively. Impacts performed with projectiles of masses $\sim$ 10 g are marked by black symbols, and $\sim$ 30 g by red symbols. The black and red lines in (a) are power law fits to these two types of impacts (10 g and 30 g projectiles), both with indexes $\sim$ 2.5. The line in (b) shows the power law fit to the entire data set, with an index of 2.5.}
 \label{f:pd_scaling}
 \end{center}
\end{figure*}

As described in the previous section, a rebound of the projectile from the target was never observed in reduced gravity conditions, compared to microgravity. This could be due to two factors: either no rebound took place, or the rebound was too small to be observed at the space and time resolution of the video recordings of the impacts. In the case that the absence of observed rebound in reduced gravity is only due to resolution limitations of the experiment cameras, we can assume that the rebound takes place with the same coefficient of restitution as in microgravity and calculate the distance and time travelled by the projectile after the impact before it gets stopped by gravity. As the only force on the projectile after it leaves the target again is gravity, simple newtonian mechanics give us the time $t_0$ at the apogee of its trajectory, and the height $h$ above the target it reaches at $t_0$:
\begin{equation}
t_0 = \frac{v_r}{g_l} \quad ; \quad h = \frac{1}{2}\frac{v_r^2}{g_l}
\end{equation}
where $v_r$ is the rebound speed, and $g_l$ the value of the local gravity acceleration.

For impact speeds around 40 cm/s, we measure coefficients of restitution around 0.02 (see Figure~\ref{f:COR}). This would set $v_r$ to 8~mm/s. Reduced gravity environments had gravity levels around 10$^{-2}g$, thus $g_l \sim$ 0.1 m/s$^2$. This leads to a travel time of $t_0$ = 80 ms and a rebound travel height of $h$ = 320 $\mu$m. During the PRIME-3 campaign, the camera resolution was around 80 $\mu$m/pixel and the recording was performed at 120 fps (about 8~ms between frames). In addition, reduded gravity environments during this campaign lead to short parabolic trajectories of the ejected particles consistent with a $\sim0.05g$ acceleration environment, leaving a clear view of the projectile during the entire recording. This means that a rebound would have been observed and a rebound motion could have been measured for 10 frames and over 4 pixels. As no rebound was observed, it appears that they were due to a different behavior of the target and no rebound actually took place in reduced gravity. To illustrate this, we show two PRIME-3 impacts with different outcomes in Figure~\ref{f:analysis}c, and their measured projectile positions, at deepest penetration (t$_0$) and after 22 (no rebound) and 4 (rebound) frames. Some impacts generating few ejecta allow for a long term trackng of the projectile after the impact, allowing for the detection of rebound speeds as low as a few mm/s. However, large ejecta blankets obstruct the view of the projectile (see Figure~\ref{f:analysis}a), thus limiting the obsevation of very low rebound speeds in these cases.

\subsection{Scaling the penetration depth}

As shown in Figure~\ref{f:pd}, scaling the maximum penetration depth of the projectile into the target ($z_{max}$) with either the impact energy or momentum was not possible. Here, the difference in gravity levels (reduced or microgravity) did not appear in the distribution of the data points. 

Following \cite{katsuragi2013}, Figure~\ref{f:pd_scaling}a shows $z_{max}$ as a function of the quantity $\mu^{-1}(\rho_p/\rho_g)^{1/2}D_p^{2/3}H^{1/3}$. In this quantity, $\mu$ is the coefficient of friction of the target material. From angle of repose measurements, we can estimate that $\mu$ = 0.84 for JSC-1 and 0.67 for quartz sand \citep{brisset2016ICE}. $\rho_p$ and $\rho_g$ are the densities of the projectile and target material, respectively; $D_p$ the projectile diameter; and $H$ the equivalent total drop distance as defined in the paragraph on penetration depth results. Figure~\ref{f:pd_scaling}a allowed us to recognize two groups of data points, which we distinguised by color: impacts with projetiles of masses around 10 g (see the PRIME-3 data in Table~\ref{t:impact_list}: 9.82 to 11.76 g) in black, and impacts with $\sim$30 g projectiles (28.2 to 30.75 g). Power law fits to each population yielded similar indexes of about 2.5, and a factor of about 3 between both. We concluded that our data set scales with $\mu^{-1}(\rho_p/\rho_g)^{0.1}D_p^{2/3}H^{1/3}$, the exponent for $\rho_p/\rho_g$ being 0.1 rather than 0.5 (all PRIME-3 projectiles have the same diameter of 2 cm). The result of this scaling is shown in Figure~\ref{f:pd_scaling}b.

This scaling and the associated power law index are in contrast with results by \cite{katsuragi2013}. In particular, the contribution of the projectile to target densities ratio is reduced from an index 0.5 to 0.25, while the projectile diameter and impact energy have increased contributions from indexes of 0.67 to 1.67 and 0.33 to 0.83, respectively. The coefficient of friction has an increased contribution from an index -1 to -2.5 ($\mu < 1$). This means that in low gravity (no distinction between reduced and microgravity), the diameter and kinetic energy of the projectiles play a more significant role in the penetration depth than the density ratio with the target material than in 1$g$. This reflects the fact that the gravity pull on the projectile during penetration is much reduced, and the density ratio between projectile and target material is a less significant factor during penetration. Large and high-velocity projectiles will penetrate the target easier in reduced gravity. The role of the cohesion between the target grains (which is implicit in the coefficient for friction $\mu$) is increased by the reduced gravity environment: cohesive targets are more efficiently stopping a projectile than in 1$g$. This supports the notion that cohesive forces can become significantly more important in low gravity compared to the weight of the target particles (see following paragraph).

\begin{figure}[t]
  \begin{center}
  \includegraphics[width = 0.5\textwidth]{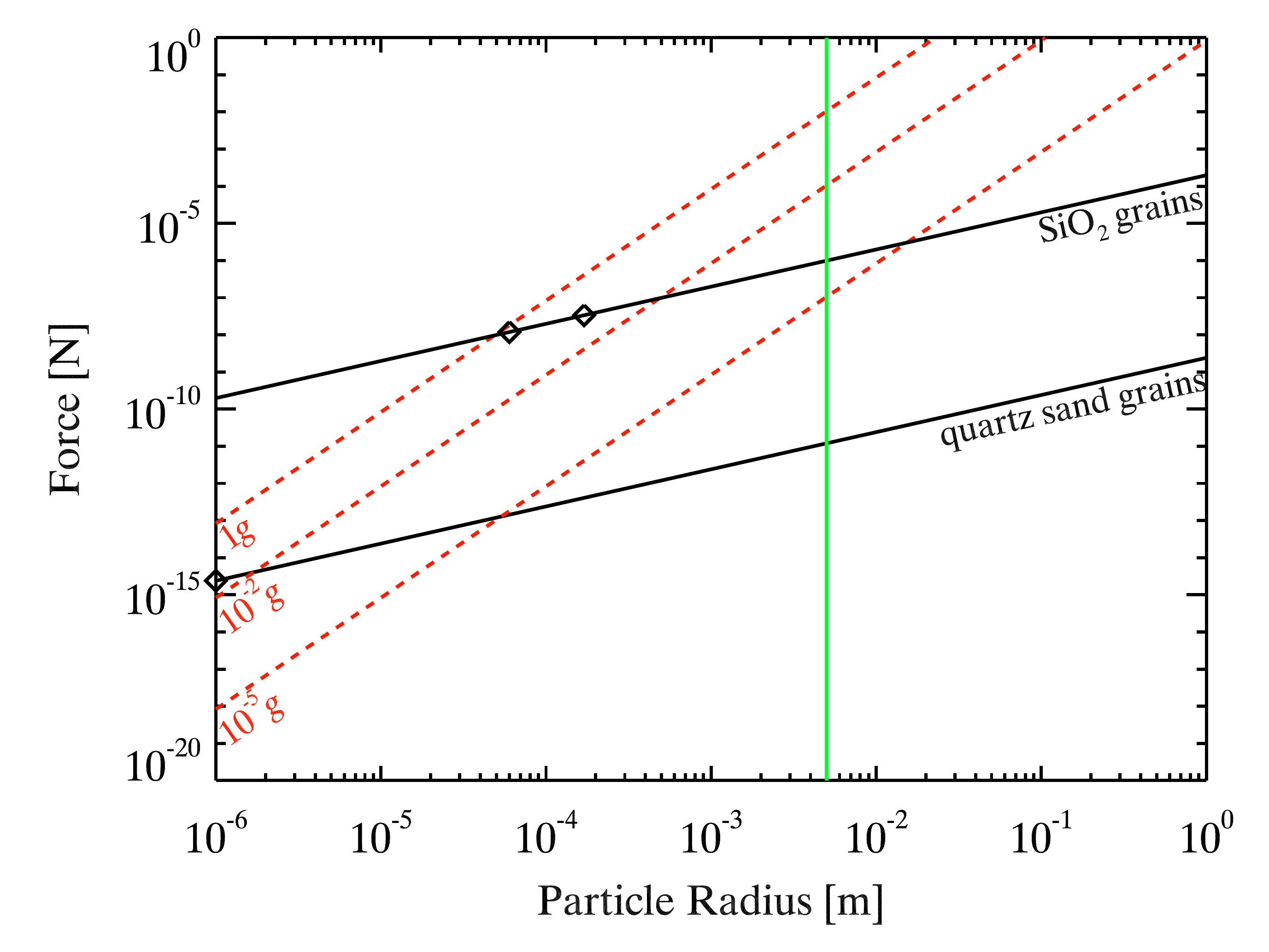}
 \caption{Comparison between gravitational and cohesive forces depending on the grain radius. The black lines correspond to the measured pull-off force between SiO$_2$ polydisperse \citep{brisset2016AA} and quartz sand grains \citep{kendall_et_al1987Nature}. The grain sizes at which the measurements were performed are marked as diamond symbols. The dashed red lines show the gravitational forces on the regolith grains for different g-level environments. The green line marks the largest particles of our target size distributions.}
 \label{f:force}
 \end{center}
\end{figure}

\subsection{Gravity-cohesion competition}

In Figure~\ref{f:force}, we compare the ambiant gravity forces to inter-particle cohesion forces (van der Waals), following the example of \cite{scheeres2010}, using measured inter-particle force values rather than model estimations. The straight black lines in this figure represent the cohesion force between two grains, which is linearly proportional to the particle radius, as measured for two different types of particles. We are using the Johnson-Kendall-Roberts theory \citep{johnson_et_al1971PRSL}, estimating the force between two grains to be the pull-off force, $F_{po}=3\pi\gamma r$, $\gamma$ being the surface energy of a grain, and $r$ its radius. The surface energies for quartz sand was measured in \cite{kendall_et_al1987Nature} and in \cite{brisset2016AA} for polydisperse SiO$_2$ particles in similar experiment conditions (vacuum levels of around 10 mTorr at room temperature). These polydisperse SiO$_2$ grains were similar to JSC-1 lunar simulant in shape, while quartz sand has more rounded, less cohesive grains. The red dashed lines mark the force of gravity on individual grains for three g-levels and the green vertical line marks the largest particles present in the target materials used in the present work. 

In this figure, we can see that, for quartz sand, the forces induced on the target grains due to gravity are several orders of magnitude larger than inter-particle adhesive forces at 1$g$ and 10$^{-2}g$. Only at 10$^{-5}g$, these two forces are of the same order for the smaller target particles. This can explain the reason for lower ejected masses and larger average ejecta speeds (at the same impact speed) observed in reduced gravity compared to microgravity: in reduced gravity, only the fastest, lighter grains can overcome the gravity field and be ejected from the target. In microgravity, the gravity force is low enough to allow for most grains to get lifted, even the slower ones, and the average ejecta speed is accordingly lower. In addition, as the gravity and cohesion forces are of the same order in microgarvity, the target material displays a more elastic behavior: cohesion forces between the particles seem to allow for a partially elastic response of the surface to the impact, thus leading to a rebound of the projectile (see Figure~\ref{f:outcomes}, open triangles and diamonds), while in reduced gravity environments the presence of a constant downward force locks the particles in place \citep[e.g. jamming, ][]{cates1998} and increases the inter-particle forces due to friction inside the material \citep{murdoch2013}, thus preventing projectile rebound. 

In Figure~\ref{f:force}, we can also see that JSC-1 targets (which approximately behave like SiO$_2$ grains) have stronger cohesion forces between grains than quartz sand: the cohesion and gravity forces are of the same order already in reduced gravity, while cohesion dominates by several orders of magnitude in microgravity. The strong dominance of the cohesive forces can explain why none of the impacts in JSC-1 in microgravity created an ejecta curtain. While most impacts in JSC-1 were performed at speeds below 40~cm/s and higher impact speeds may have possibly triggered larger ejecta masses, impacts at similar speeds and in the same micrograivty environment generated an ejecta curtain with sand target (Figure~\ref{f:mass}). The predominance of cohesive forces between JSC-1 grains kept target material from escaping. 

We can also see a difference in behavior of the JSC-1 targets in reduced- and microgravity environments: for the same impact velocities, rebound is observed in microgravity, while the projectile embeds into the target in reduced gravity (filled black diamonds and red circles in Figure~\ref{f:outcomes}). Even though JSC-1 displays a more cohesive behavior than sand in reduced gravity, the data inidicates that these cohesion forces do not dominate gravity enough at 10$^{-2}g$ to lead to the rebound of the projectile.

\subsection{Interacting with small body surfaces}

The experiments presented here are limited by the size of the target tray and projectile. However, it is possible to learn from their results about possible outcomes of surface interactions by spacecraft landers or sampling mechanisms. In particular, impact speeds under 1~m/s are typical for past and planned missions interacting with the surface of a small body; Philae touched down on the surface of 67P/Churyumov-Gerasimenko for the first time at $\sim$ 1 m/s, OSIRIS-REx's Touch-And-Go (TAG) mechanism is planned to impact the surface of Bennu at around 20 cm/s, Hayabusa-2's MASCOT will land on Ryugu also at several 10 cm/s. The present experiments are therefore in a relevant impact speed regime. While the regolith size distribution on asteroid surfaces seems to heterogenously include large and fine particles, images of the surfaces of asteroids visited by spacecraft show the presence of regions of very smooth terrain covered in grains of sizes below a mm \citep[e.g. the smooth ponds on Eros;][]{cheng2002}. These smooth terrains also represent ideal surface interaction sites as they reduce the risks related to landing on uneven surfaces. Therefore, the target size distributions of our experiments are relevant to understanding spacecraft interactions with asteroid surfaces. 

At these relevant impact speeds and grain sizes, we observed the new phenomenon of projectile bouncing off of granular material surfaces in microgravity. While numerical simulations are currently the only way of studying low-velocity impacts of spacecraft landers at the correct size scales, those simulations are mostly using hyper-velocity or 1$g$ experiment impact data for calibration. The only simulations predicting and reproducing projectile bouncing from the surface \citep{maurel2017}, for example, are using Earth gravity impact experiments at $>$10 m/s into granular material \citep{yamamoto2006} or impacts on hard surfaces as input parameters \citep{biele2017}. \cite{maurel2017} specifically indicate the need for parameter refinement based on microgavity data from experiments.

It can be noted that the experimental conditions, in particular the level of residual accelerations, have a significant impact on the target behavior. Even for irregular grains that behave like cohesive powders, the granular material displayed a different behavior in reduced- and microgravity. This means that the behavior of granular material at the g-levels present on the surface of small asteroids (10$^{-4}g$ and below) cannot be extrapolated from its behavior at Lunar g-level ($\sim10^{-2}g$). Therefore, experiments relevant to the interaction with surfaces of bodies like Ryugu and Bennu will require the high-quality microgravity environment of free-fall drop towers or free-floating experiment hardware.

\section{Conclusion}

We have combined and analyzed low-velocity impacts into regolith from five experimental campaigns. The PRIME and COLLIDE experiment setups allowed for recording impacts of cm-sized spherical projectiles into a bed of regolith simulant (quartz sand, JSC-1 Lunar, and JSC Mars-1 simulant), sieved to sizes under 250~$\mu$m. Due to their flight platform, these experiments were run either at reduced gravity levels of $\sim10^{-2}g$ when fixed to the parabolic aircraft or during 0.05$g$ parabolas, or at microgravity levels of $<10^{-4}g$, when free-floating in the airplane or during Shuttle and suborbital rocket campaigns. The impacts into regolith simulant resulted into four types of outcomes, some producing ejecta and some displaying projectile rebound. These results revealed major differences in the target behavior depending on if the experiment was run in reduced- or microgravity conditions. The main results of the data analysis can be summarized as follows:
\begin{itemize}
\item We observed projectile rebound off the target in microgravity. None of the impacts in reduced gravity displayed a similar behavior. The coefficient of restitution of the impacts varied with the impact energy following a power law of index -1/4 (Figure~\ref{f:COR});
\item The maximum penetration depth was observed to depend more on the projectile size and energy and less on its density difference with the target than in 1$g$;
\item For the same impact speeds, more mass was ejected in microgravity compared to reduced gravity, but the average ejecta speed was lower (Figures \ref{f:ejecta} and \ref{f:mass});
\item The difference in cohesive forces in quartz sand and JSC-1 could be observed from the experiment results, with quartz sand behaving like a cohesionless material and JSC-1 displaying cohesive behavior. None of the impacts into JSC-1 generated an ejecta curtain in microgravity, and most of them displayed projectile rebound.
\end{itemize}

The results of these flight campaigns show that the acceleration environment at which experiments are performed have a significant influence on their outcome. The study of regolith behavior at the surface of small bodies ($<$ 1~km-sized asteroids) will require acceleration environments below 10$^{-4}g$, as data cannot be extrapolated from behavior in reduced gravity environments ($>10^{-3}g$).

From our current data set, it clearly appears that more low-velocity impact data is required in microgravity conditions. This is due to the fact that clean microgravity environments are difficult to obtain, with most platforms retaining residual acceleration levels around 10$^{-2}g$. In addition, more data is currently available for sand targets than for JSC-1 lunar simulant. As JSC-1 is more representative of the granular material expected on asteroids (in particular due to the irregular particle shape), data on JSC-1 targets is more relevant for application to small body surfaces. We can also note that all experiments presented here were run using similar target size distribution: particles sieved to sizes $< 250 \mu$m. From images returned of asteroid Itokawa \citep{fujiwara2006} and spectral analysis of Ryugu, we can infer that some asteroid surface regolith has larger average grain sizes. For a future contribution of our experimental results into small body surface exploration missions, the investigation of grain behavior at sizes between 1 mm and 1 cm will be necessary. Such future experiments will be even more sensitive to the quality of the microgravity environment they will be run in due to the larger grain sizes (see Figure~\ref{f:force}). 

\section*{Abbreviations}

\noindent CMR: Center for Microgravity Research; \\
COLLIDE: COLLisions Into Dust Experiment;\\
IBS: Impactor Box System;\\
ISRU: In-Situ Resource Utilization;\\
JSC: Johnson Space Center;\\
MASCOT: Mobile Asteroid Surface Scout;\\
NASA: National Aeronautics and Space Administration;\\
OSIRIS-REx: Origins, Spectral Interpretation, Resource Identification, Security, Regolith Explorer;\\
PRIME: Physics of Regolith Impacts in Microgravity Experiment;\\
TAG: Touch-And-Go.

\section*{Competing interests}

\noindent N/A

\section*{Funding}

\noindent This research is based in part upon work supported by the National Aeronautics and Space Administration under Grant No. NNX11AQ87G issued by the Planetary Geology and Geophysics Program and under grant NNX12AK43G issued by the Outer Planets Research Program, by the NASA Flight Opportunities Program and by funding from Space Florida for the Center for Microgravity Research. 

\section*{Authors' contributions}

\noindent JC secured funding for all performed experiments. JC provided the COLLIDE-1 and -2, and PRIME-1 data. JC, AD, and JB performed the experimental data collection during the PRIME-3 and COLLIDE-3 campaigns.\\
JB, SA, CC, and NM worked on the data analysis.

\section*{Authors' information}

\noindent Authors can be contacted at the following email addresses: JB: julie.brisset@ucf.edu; JC: jec@ucf.edu; \\
AD: adove@ucf.edu; SA: sumayya@knights.ucf.edu; CC: tylercox@knights.ucf.edu; NM: mohan10@knights.ucf.edu

\section*{Availability of Data and Materials}
\noindent Please contact author for data requests.

\section*{Acknowledgement}
\noindent This research is based in part upon work supported by the National Aeronautics and Space Administration under Grant No. NNX11AQ87G issued by the Planetary Geology and Geophysics Program and under grant NNX12AK43G issued by the Outer Planets Research Program, by the NASA Flight Opportunities Program and by funding from Space Florida for the Center for Microgravity Research.

\noindent We would like to thank our two anonymous reviewers for their valuable input and comments.\\



\end{document}